\documentclass[11pt,a4paper]{article}
\usepackage{jcappub}
\usepackage[utf8]{inputenc}
\usepackage[table]{xcolor}
\usepackage{verbatim}
\usepackage{amsmath,amsfonts,amssymb}
\usepackage{subfloat}
\usepackage{float}
\usepackage{xcolor}
\usepackage{array, booktabs, tabularx, makecell, multirow}
\usepackage{multicol}
\usepackage{multirow}
\usepackage{tikz}
\usetikzlibrary{shapes.geometric, arrows}
\usepackage{tabularx}
\newcolumntype{s}{>{\hsize=.5\hsize}X}

\usepackage[misc]{ifsym} 

\tikzstyle{startstop} = [rectangle, rounded corners, minimum width=3cm, minimum height=1cm,text centered, draw=black, fill=red!30]
\tikzstyle{io} = [trapezium, trapezium left angle=70, trapezium right angle=110, minimum width=3cm, minimum height=1cm, text centered, draw=black, fill=blue!30]
\tikzstyle{process} = [rectangle, minimum width=3cm, minimum height=1cm, text centered, text width=3cm, draw=black, fill=orange!30]
\tikzstyle{decision} = [diamond, minimum width=3cm, minimum height=1cm, text centered, draw=black, fill=green!30]
\tikzstyle{arrow} = [thick,->,>=stealth]

\tikzstyle{lcdmparam} = [rectangle, rounded corners, minimum width=3cm, minimum height=1cm,text centered, draw=black, fill=yellow!30]
\tikzstyle{stockcode} = [rectangle, rounded corners, minimum width=3cm, minimum height=1cm,text centered, draw=black, fill=red!30]
\tikzstyle{modcode} = [rectangle, rounded corners, minimum width=3cm, minimum height=1cm,text centered, draw=black, fill=blue!30]
\tikzstyle{modparam} = [rectangle, rounded corners, minimum width=3cm, minimum height=1cm,text centered, draw=black, fill=orange!30]
\tikzstyle{goal} = [rectangle, rounded corners, minimum width=3cm, minimum height=1cm,text centered, draw=black, fill=green!50]

\usetikzlibrary{mindmap,backgrounds}
\usepackage{adjustbox}
\usepackage{tabularx}
\usepackage[normalem]{ulem}
\hypersetup{%
,urlcolor=red
,citecolor=red
,linkcolor=blue
}

\def\mnras{Monthly Notices of the Royal Astronomical Society}
\def\aap{Astronomy and Astrophysics}

\def\jcap{Journal of Cosmology and Astroparticle Physics}
\def\apj{Astrophysical Journal}
\def\physrep{Physical Reports}
\def\prd{Physical Review D}

\author[a]{Sankarshana Srinivasan,}
\author[b]{Daniel B Thomas,}
\author[c, d, e]{and Peter L. Taylor}

\affiliation[a]{Universit\"{a}ts-Sternwarte, Fakult\"{a}t f\"{u}r Physik, Ludwig-Maximilians Universit\"{a}t, Scheinerstraße 1, 81679 M\"{u}nchen, Germany}
\affiliation[b]{Jodrell Bank Centre for Astrophysics, Department of Physics and Astronomy, University of Manchester, Oxford Road, Manchester, M13 9PL, United Kingdom}
\affiliation[c]{Center for Cosmology and AstroParticle Physics (CCAPP),
The Ohio State University, Columbus, OH 43210, USA}
\affiliation[d]{Department of Physics, The Ohio State University, Columbus, OH 43210, USA}
\affiliation[e]{Department of Astronomy, The Ohio State University, Columbus, OH 43210, USA}

\emailAdd{ssriniva@usm.lmu.de}
\emailAdd{dan.b.thomas1@gmail.com}
\emailAdd{taylor.4264@osu.edu}

\date{April 2020}

\abstract{
Stage IV large scale structure surveys are promising probes of gravity on cosmological scales. Due to the vast model-space in the modified gravity literature, model-independent parameterisations represent useful and scalable ways to test extensions of $\Lambda$CDM. In this work we use a recently validated approach of computing the non-linear $3\times 2$pt observables in modified gravity models with a time-varying effective gravitational constant $\mu$ and a gravitational slip $\eta$ that is binned in redshift to produce Fisher forecasts for an LSST Y10-like survey. We also include in our modelling an effective nulling scheme for weak-lensing by applying the BNT transformation that localises the weak-lensing kernel enabling well-informed scale cuts.  We show that the combination of improved non-linear modelling and better control of the scales that are modelled/cut yields high precision constraints on the cosmological and modified gravity parameters. We find that 4 redshift bins for $\mu$ of width corresponding to equal incremental $\Lambda$CDM growth is optimal given the state-of-the-art modelling and show how the BNT transformation can be used to mitigate the impact of small-scale systematic effects, such as baryonic feedback.  
 }


\title{Cosmological gravity on all scales IV: 3x2pt Fisher forecasts for pixelised phenomenological modified gravity}

\begin{document}

\maketitle

\section{Introduction}

Over the past few decades, the $\Lambda$CDM model has emerged to be successful in describing various observational data-sets. In particular, measurements of the cosmic microwave background (CMB) \cite{Planck2018_VIII} and the large scale structure of the universe \cite{Abbott_2022, more2023hypersuprimecamyear3, miyatake2023hypersuprimecamyear3, Sugiyama_2023, Asgari_2021, Alam_2021} have measured the 6 parameters of $\Lambda$CDM to within percent levels of precision. As cosmology enters into an era of unprecedented precision and data volumes going forward into the era of stage IV surveys, the fundamental nature of the main ingredients of $\Lambda$CDM, i.e., the cosmological constant $\Lambda$ and the cold dark matter (CDM) remain unknown. Efforts to compute the value of $\Lambda$ in quantum field theory from the zero-point energy of the vacuum yields a number that is discrepant with the observed value by over 50 orders of magnitude, popularly referred to as the cosmological constant problem \cite{MartinCC2012}. On the other hand, the standard hypothesis that cold dark matter consists of a particle that couples to the standard model very weakly has not yet yielded any detections, although many experiments are active and taking data around the world.

The assumption that General Relativity (GR) is valid across all cosmological scales is crucial in $\Lambda$CDM, and to the inference that $\Lambda$ and CDM have the properties that they are believed to have. Therefore, a large model-space of modified gravity (MG) models \cite{ref:CliftonReview, Nojiri_2017, Ezquiaga_2018, saridakis2023modified} has been created to explore this possibility. In order to efficiently test this model space, there have been a number of successful MG parameterisations in which one can test a large class of these models. On local, highly non-linear scales, the Newtonian limit of modified gravity models has been used in the so-called parameterised post-Newtonian (PPN) method with great success in establishing the validity of GR to a high precision (\cite{ref:WillGrav}). While these scales are typically not of interest to cosmologists, it is important to note that any MG phenomenology has to recover GR on solar-system scales. Some models ensure that they satisfy the Solar System data via so-called screening mechanisms and a number of such mechanisms exist in the literature to allow MG models to evade such constraints \cite{Brax_2021}. However, we note that this is not required to satisfy the Solar System data; e.g. several classes of theories that do not require screening are those covered by the extension of the PPN formalism to cosmological scales \cite{Sanghai_Clifton_2016, 2017Sanghai}. Using this extension, the PPN parameters have recently been constrained using cosmological data \cite{ThomasPPNC2} data.

Progress on testing gravity on cosmological scales has been made via the so-called quasi-static approximation, i.e., the small-field low-velocity limit in linear perturbation theory (See \cite{ref:Gleyzes, ref:BattyePearson} for examples). As a result, most of the cosmological tests and the constraints obtained on MG parameters apply only on linear scales, i.e., where the density contrast $\delta \ll 1$. Thus, there are clearly two regimes where gravity is well-tested, i.e., solar-system scales and linear cosmological scales. Assuming the validity of GR across all scales involves an extrapolation of many orders of magnitude over all the scales in between these two regimes. 

Non-linear cosmological structure formation is an important range of scales between these two regimes, and is a regime where the assumption that GR applies has not yet been tested. It can be quite complicated to model this non-linear structure formation even in $\Lambda$CDM, requiring computationally expensive $N$-body simulations with varying levels of complexity in the hydrodynamic astrophysical modelling (see for example \cite{Hernandez_Aguayo_2023}). More recently, there has been a lack of convergence across these simulations in the matter power spectrum $P(k)$, commonly attributed to baryonic feedback (see \cite{Chisari_2019, Kids_Baryons_SZ}), i.e., energy injection from astrophysical phenomena. Therefore, a crucial challenge going forward is to understand the range of scales in large scale structure that can be reliably modelled in order to test cosmology. This is particularly important because a large fraction of the data that will be taken in upcoming surveys, such as Euclid and LSST, will be on these scales. 

In $\Lambda$CDM, a number of techniques exist to push as deep into the non-linear regime as possible while remaining robust in the modelling (see for example \cite{Baumann_2012, Zhen_HOD, 2ptcollaboration2024parametermaskedmockdatachallenge}. Unfortunately, such techniques do not exist in modified gravity, beyond a handful of specific cases that model f(R) or DGP gravity. This has led to severe scale cuts being employed, an example being in the beyond $\Lambda$CDM analysis of the Dark Energy Survey \cite{DES_2019_extensions}, which seriously limits constraining power. 

Recently, it was shown \cite{ref:DanPF} that modified gravity can be parameterised on all cosmological scales via the so-called post-Friedmann formalism. Modified gravity phenomenology appears in this parameterisation via a re-scaling of the Poisson equation in GR (quantified by the parameter $\mu$) and a re-scaling of the ratio of the gravitational potentials in GR (quantified by the parameter $\eta$). We use this approach and take a model-agnostic approach where we bin the modified gravity parameters in redshift in order to avoid making modelling assumptions such as a specific functional form for the Lagrangian or field-content. In previous work \cite{Srinivasan2021, Srinivasan_2024}, we ran $N$-body simulations with the modified gravity parameters binned in redshift and showed that it is possible to predict $P(k)$ in our simulations with a modified version of the \texttt{ReACT} \cite{BoseReACT} formalism. In this work, we will forecast the constraining power of an LSST-like survey on the MG parameters $\mu$ and $\eta$. In particular, we will aim to answer the following questions: 

\begin{itemize}
    \item Can the BNT transform (or equivalent nulling schemes) be used to mitigate uncertainties from the small scales such as baryonic feedback? 
    \item What is the most efficient binning scheme in order to optimise constraining power? 
    \item How does the constraining power scale with the scale-cut employed and non-linear modelling choices? 
    \item What are the relevant degeneracies between parameters (MG or cosmological) on the range of scales we can model the matter power spectrum? Do these get affected by the choice of parameterisation? 
\end{itemize}

Note that there are a number of works in the literature in which Fisher forecasts of phenomenological modified gravity are used to quantify the increase in constraining for stage IV surveys \cite{ref:Casas2017, Aparicio_Resco_2020, Casas_2023}. A common feature in all of them is that they assume specific functional forms for $\mu(z)$ and $\eta(z)$ that come from known models/phenomenology instead of the binning approach adopted here which is model-agnostic. An exception is \cite{ref:Casas2017}, where the authors consider five equal size bins of width $\Delta z = 0.5$ between $0\leq z \leq 2.5$, with the Hu-Sawicki PPF framework being adopted for the non-linear $P(k)$. The authors in this paper also make use of a principal component analysis to compute the bin-combination that is best constrained by the data. While this is a useful and relatively common technique of re-parameterisation to understand the most constrained combination of model parameters, we believe that it is premature to apply such techniques when the non-linear modelling has not been formally validated for multiple $\mu(z)$ bins. The same hurdle also prevents a direct reconstruction of $\mu(z)$ and $\eta(z)$ from the data using Gaussian processes, as done in \cite{Calder_n_2023}. Crucially, the approach taken in this work is fully validated against N-body simulations.

We will begin in sec.~\ref{sec2} by describing our modified gravity formalism in sec.~\ref{sec2.1} and also briefly review our implementation of the ReACT code sec.~\ref{sec2.2}. In sec.~\ref{sec2.3}, we describe the method we use to make our scale cuts for our modified gravity Fisher forecast using the BNT transformation. We then define a pipeline developed using the modular software \texttt{Cosmosis} \cite{ref:ZuntzCosmosis} to model an LSST-like survey and perform Fisher forecasts on a simulated $3\times 2$pt data vector. We present our results and contextualise them in sec.~\ref{sec3}. We then summarise and conclude in sec.~\ref{sec4}.

\section{Methodology}\label{sec2}

In this section, we review the key ingredients in our modelling of cosmological observables relevant to the $3\times 2$pt data vector we construct in order to perform Fisher forecasts. We begin with a discussion on the modified gravity formalism and the modifications we have made to the \texttt{ReACT} library to compute the non-linear matter power spectrum in our parameterisations.  

\subsection{Binning the Poisson equation: the Post-Friedmann formalism}\label{sec2.1}

We make use of the post-Friedmann formalism that was introduced in \cite{ref:Milillo} and developed to include non-linear modified gravity in \cite{ref:DanPF}, focusing on the ``Parameterised Simple 1st Post-Friedmann'' (PS1PF) equations developed in \cite{ref:DanPF}. In this approach, the modifications to gravity relevant on cosmological scales are captured via the modifications to the Poisson equation and the gravitational slip as shown below
\begin{eqnarray}
\frac{1}{c^2}k^2 \tilde{\phi}_{\rm P} & = & -\frac{1}{c^2}4\pi a^2 \bar{\rho} G_{\rm N}\mu(a,k)\tilde{\Delta} \,, \label{eq:MGParam} \\
\tilde{\psi}_{\rm P} & = & \eta(a,k)\tilde{\phi}_{\rm P} \, ,
 \end{eqnarray}
where $\bar{\rho}$ is the background density, $\tilde{\Delta} = \tilde{\delta} - \frac{\dot{a}}{a}\frac{3}{c^2k^2}ik_i \tilde{v}_i$ is the gauge-invariant density contrast in Fourier space (which crucially is not required to be small in the PS1PF approach), $G_{\rm N}$ is Newton's constant, $\phi_P$ and $\psi_P$ are the standard Newtonian gravitational potentials\footnote{$g_{00}=-1-\frac{2\phi_P}{c^2}$; $g_{ij}=a^2\delta_{ij}\left(1-\frac{2\psi_P}{c^2} \right)$.} (normally found to be equal in GR) and $\mu(a, k)$ is a dimensionless function of space and time representing a change to the strength of gravity. These equations are derived by expanding the FLRW metric to one step beyond the leading order, at which order the equations describe structure formation on all scales (see \cite{ref:DanPF} for details). One of the key consequences of this parameterisation is that the dynamics of massive particles are purely governed by $\mu(a, k)$, which can be probed in dark matter only $N$-body simulations. The second parameter $\eta$, the so-called slip parameter purely affects photon geodesics and, therefore, can be modelled in post-processing. It was shown in \cite{ref:DanPF} that this parameterisation may be adapted to any modified gravity model with a well-defined Newtonian limit and a sufficiently small vector potential on cosmological scales. From a PS1PF perspective, the work in \cite{ref:Hassani2019, ref:HassaniNBodyMG} can be interpreted as showing that specific choices in terms of models and screening mechanisms may be mapped on to specific functional forms for $\mu$ in this approach. 

Our philosophy here is to focus on searching for as generic deviations from GR as possible, and in the process perform a null test of the GR$+\Lambda$CDM paradigm. Such null tests are important to carry out in the upcoming era of precision of cosmology, as one is being confronted with various tensions with $\Lambda$CDM.   With these points in mind, we have decided to take as model-agnostic an approach as feasibly possible by binning $\mu$ and $\eta$ in redshift. We note that this restriction to time dependence is a choice that we are making for this forecast, rather than a requirement of the underlying framework that we are using. We have shown \cite{Srinivasan2021, Srinivasan_2024} that there is already rich phenomenology in the purely time-dependent binned case that warrants a dedicated analysis before one introduces additional complexity in the form of scale-dependence.

In this work, we adopt the same binning scheme as we used in our previous two papers where we implemented piece-wise constant pixels in redshift for $\mu$. We only `switch on' modified gravity, i.e. vary $\mu$ or $\eta$ from their GR value, in one pixel at a time. This essentially amounts to setting the fiducial value of $\mu=1$ in all bins when we carry out our Fisher forecasts. Our bin widths are determined by the change in the linear growth factor $D(z)$ (evaluated in $\Lambda$CDM) between the start and end of a redshift bin, i.e., the incremental growth. There are two reasons for this choice, the first being this can be computed analytically in linear theory. The second is that this choice naturally modulates our bin widths such that they allow for more complex structure formation at late times compared to early times. In other words, our redshift resolution in $\mu(z)$ increases as non-linearity starts to become more important. As such, we continue to work with redshift bins of constant incremental $D(z)$ in $\Lambda$CDM. We use the publicly available Python wrapper of the \texttt{isitGR} Boltzmann solver \cite{Dossett_2011, Dossett_2012, Garcia_Quintero_2019} to obtain the linear matter power spectrum in our pipeline and a modified version of the \texttt{ReACT} code for the non-linear matter power spectrum. While the effects of $\eta \neq 1$ are not captured by $P(k)$, they are captured in the weak-lensing convergence power spectrum. We will return to this point later in the paper when we discuss our results.

\subsection{Computing the non-linear matter power spectrum with \texttt{ReACT}}\label{sec2.2}

The principle behind the halo model reaction formalism \cite{ref:ReactTheory, Cataneo_2019, BoseReACT, Bose_2021} is that one can greatly improve the accuracy in the halo model in the transition regime between the 2-halo term and the 1-halo term by computing the ratio dubbed the `non-linear response of the power spectrum' or `reaction' \cite{Mead2017} i.e., the ratio of matter power spectra between two cosmologies that share the same linear power spectrum. The halo model reaction formalism computes the non-linear reaction for beyond $\Lambda$CDM cosmologies, given by the ratio of $P(k)$ in the target cosmology relative to the `pseudo spectrum', which is the $\Lambda$CDM power spectrum with the the initial conditions modified to match the linear power spectrum in the modified gravity model at a specific target redshift. The code that implements this formalism, \texttt{ReACT} has been applied to phenomenological extensions of $\Lambda$CDM with specific predictions for the functional form of $\mu(z, k)$ \cite{Bose_2023, MGCAMB_React}. In the case where theoretical predictions for $\mu$ are derived from the effective field theory of dark energy, the non-linear recipe in \texttt{ReACT} was validated against emulators for $w$CDM, $f(R)$, $n$DGP cosmologies to within sub-percent accuracy. 

\begin{figure}
    \centering
    \includegraphics[width = 0.7\textwidth]{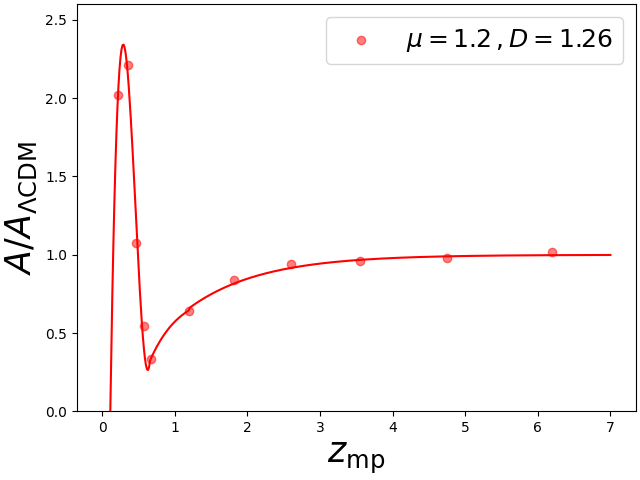}
    \caption{The concentration fitting function at $z=0$ for $D(z) = 1.26$ and $\mu=1.2$. As is clear in this plot, the significant effect on the halo concentration is when $\mu$ is modified at late times. This will prove to be important later when we present our results.}
    \label{fig:conc_fit}
\end{figure}

In our parameterisation, the Poisson equation is derived making no distinction between the clustering of matter due to gravity on linear vs non-linear scales (see sec. IV A in \cite{ref:DanPF}). Therefore, in order to make contact with our simulations within the the \texttt{ReACT} framework, one sets the spherical collapse parameter $\mathcal{F} \equiv \mu(a, k)-1$. This is our basic non-linear recipe. In addition to this recipe, we modified the \texttt{ReACT} code to compute the non-linear matter power spectrum in $N$-body simulations with $\mu$ modified in piece-wise constant bins at $z\leq 7.$ to within 1\%. This was achieved by implementing a fitting function for the ratio of the amplitude of the  concentration-mass relation in modified gravity relative to the pseudo cosmology as a function of $\mu(z)$ in the \texttt{ReACT} code for $0.8 \leq \mu \leq 1.2$ and $1.26 \leq D(z) \leq 1.45$. The general functional form for the fit is given by
\begin{equation}\label{eq;MainFit}
    A_{\rm MG}/A_{\rm pseudo} = \begin{cases}
    f_1(\bar{z}) & \text{if } \bar{z}\geq 1.2\\
    f_2(\bar{z}),& \text{if } \bar{z}\leq 1.2  
\end{cases}
\end{equation}
where $\bar{z} = z_{\rm mp} - z_{\rm pk}$ and $z_{\rm pk}$ is the redshift at which the power spectrum is computed. We divide the redshift range into two regimes. In the first regime, i.e., $1.2\leq \bar{z} \leq 7$,  we use an exponential function 
\begin{eqnarray}
    f_1(\bar{z}) = C_1e^{-\bar{z}} + 1\, ,
\end{eqnarray}
 where $C_1<0$ when $\mu>1$ (and vice-versa for $\mu<1$). This form ties back to the qualitative prediction that we made from hierarchical structure formation, i.e., modifying $\mu$ at later times leads to larger halos and therefore smaller concentrations relative $\Lambda$CDM \cite{ref:MultiDark}. We fit the the behaviour of $A/A_{\rm \Lambda CDM}$ at very late redshifts at $z_{\rm mp}<1.2$ using a cubic polynomial given by
\begin{equation}
    f_2(z) = (C_2 + C_3\bar{z} + C_4\bar{z}^2 + C_5\bar{z}^3) \, .
\end{equation}
The interested reader may refer to the appendix in \cite{Srinivasan_2024} to see how $C_i = \{C_1, C_2, C_3, C_4, C_5\}$ vary as a function of $\{\mu, D\}$. We show in fig.~\ref{fig:conc_fit} what the fitting function looks like for the modelling choice we employ in this paper, i.e., $D(z)$ = 1.26, the binning is shown in table \ref{tab:bins}. This choice ensures that we have good redshift resolution while still retaining the ability to accurately model $P(k)$.

\begin{table}
    \centering
 
    \begin{tabular}{|c|c|}
       \hline
       Bin 1 & $0 \leq z \leq 0.43$ \\
       Bin 2 & $0.43 \leq z \leq 0.91$\\
       Bin 3 & $0.91 \leq z \leq 1.47$\\
       Bin 4 & $1.47 \leq z \leq 2.15$\\
       Bin 5 & $2.15 \leq z \leq 3.0$\\
       \hline
    \end{tabular}

    \caption{The redshift bins for $\mu(z)$ and $\eta(z)$ in this work. We show in sec.~\ref{subsec:binning_schemes} that the bin 5 is actually very poorly constrained by the data and thus, in our baseline analysis, we only consider the first four bins. }
    \label{tab:bins}
\end{table}

The concentration-mass relation is quite difficult to measure directly from observations and exhibits significant scatter even when measured from simulations \cite{ref:MultiDark}. However, one would expect that modifying $\mu$ should impact the growth history of the Universe and as we will show in this paper, taking this fact into account might have important consequences on the constraining power in large scale structure surveys.

\subsection{Scale cuts with the BNT transformation}\label{sec2.3}
As mentioned previously, one of the biggest challenges in cosmology today and indeed in the coming decade is to mitigate the effects of baryonic feedback and other systematic effects relevant on cosmological non-linear scales. The current way of dealing with the small scales is to remove them entirely from the analysis, particularly in the case of extensions to $\Lambda$CDM. While this is an effective method to ensure that the analysis is trustworthy, this is no longer justifiable going forward towards stage IV surveys due to the severe loss of constraining power. We will explicitly show this in sec.~\ref{sec3}.

\begin{figure}
    \centering
    \includegraphics[width = 0.45\textwidth]{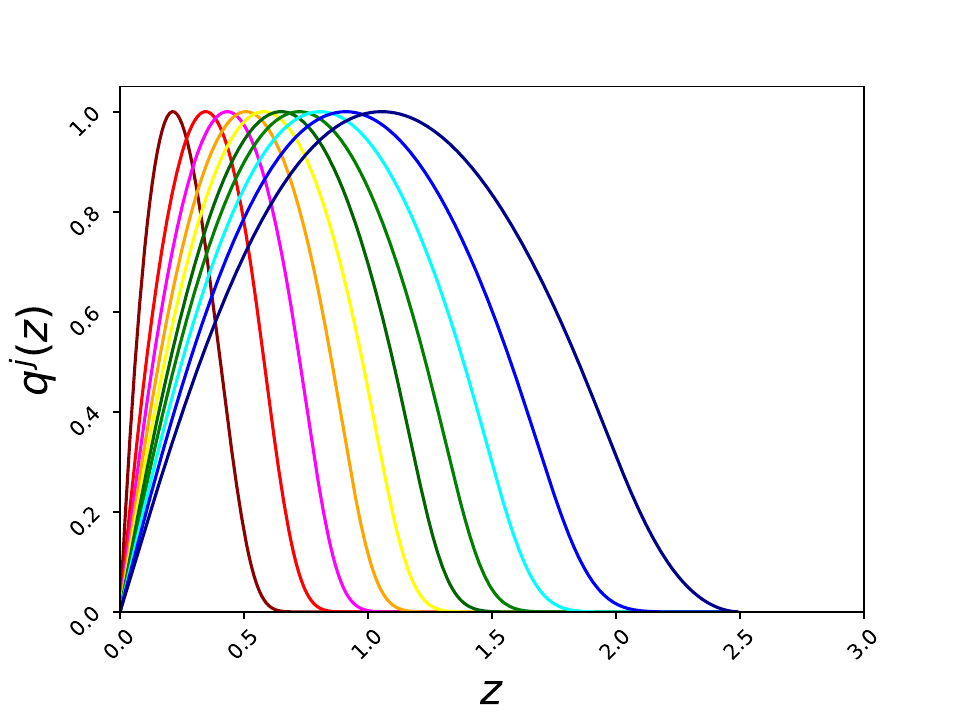}
    \includegraphics[width = 0.45\textwidth]{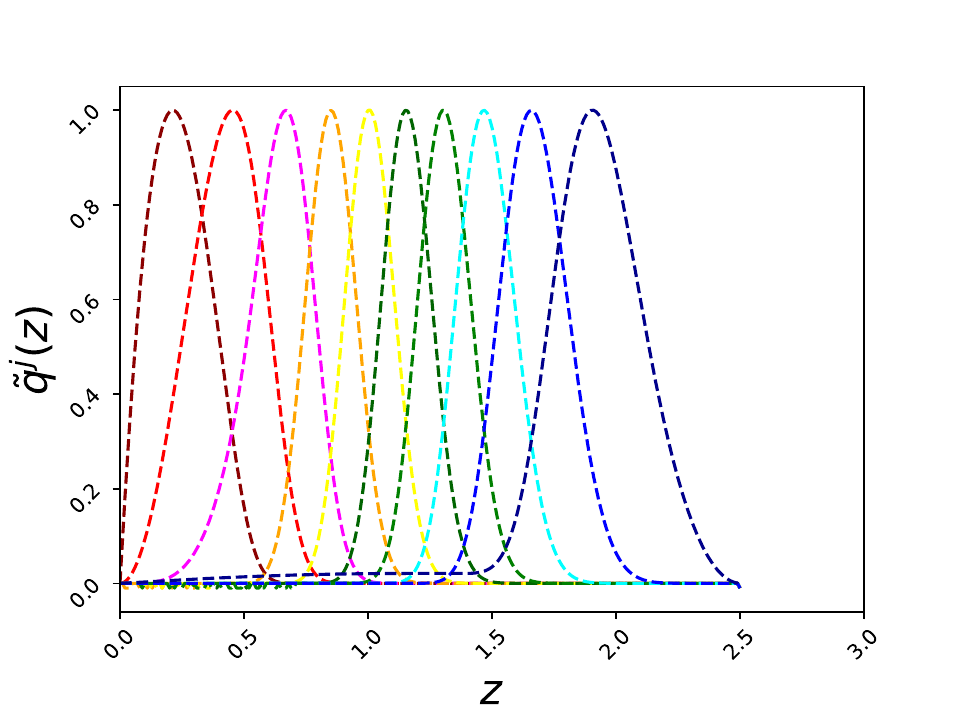}
    \caption{The left panel shows the lensing efficiencies in a Euclid-like survey with 10 tomographic bins and on the right panel, the same kernels after being BNT-transformed. Due to the fact that the transformed lensing kernels are narrower and overlap less, each kernel can be associated to a range of physical scales, which allows a more informed scale cut.  }
    \label{fig:lensing}
\end{figure}

In order to address this problem, we make use of the $k$-cut cosmic shear technique \cite{Taylor_2018, Taylor_2021} \footnote{https://github.com/pltaylor16/x-cut}. This method employs the Bernardeau-Nimishi-Taruya (BNT) transformation to linearly re-weight (thereby preserving information content) the lensing kernels in such a way as to ensure compactness in redshift. Mathematically, this may be achieved by choosing the weights $p_i$ so the lensing kernels $W_i$ are bounded in redshift \cite{Bernardeau_2014}. This allows one to relate each kernel to a range of comoving scales between the boundaries of the lensing kernels $\chi_{\rm min} \leq \chi \leq \chi_{\rm max}$, with the weighted average being the `typical scale' associated to each bin. Crucially, one can then relate the scale cuts applied to the lensing data to a physical scale $k_{\rm cut}$ allowing scale cuts to be made based on the modelling of the suppression of $P(k)$ from baryonic feedback.

This re-weighting is shown in fig.~\ref{fig:lensing} for tomographic bins in Euclid-like survey. It has been shown that this re-weighting increases the range of scales where the 2-halo term, attributed to large scales, contributes to the lensing signal, relative to the 1-halo term which is attributed to the harder to model non-linear scales. The BNT transformation has also been shown to have very limited dependence on the background cosmology \cite{Taylor_2021, euclid_kcut} and due to it being linear, can be inverted to obtain the original lensing $C(\ell)$. The BNT transformation has been used to apply scale cuts to constrain $f(R)$ gravity using the Subaru HyperSuprime-Cam (HSC) data \cite{Vazsonyi_2021}.

In this paper, we impose scale cuts on the the various BNT transformed $C_{ij}(\ell)$ by cutting above all $\ell_{\rm max}$ that correspond to scales below $k_{\rm max}\chi_{\rm avg}$, where $\chi_{\rm avg}$ is the minimum average distance (where the average is over the lensing kernel $q(z)$ and not $n(z)$) for the relevant tomographic bin combination, i.e., $$\chi_{\rm avg} = \text{min}\left(\chi^i_{\rm avg}, \chi^j_{\rm, avg}\right)\, .$$ We will present our results for three different choices for $k_{\rm max}$ that capture the extent to which one trusts the modelling of the non-linear scales in the analysis. For the un-transformed $C_{ij}(\ell)$, we will use the same criteria for the galaxy clustering and galaxy-galaxy-lensing data, but choose appropriate values of $\ell_{\rm max}$ for the shear data. We compute the covariance matrix for the $k$-cut statistics included in the $3\times 2$pt data vector via the likelihood sampling method described in section III E of \cite{Taylor_2021}. 

\begin{figure}
    \centering
    \includegraphics[width=0.5\linewidth]{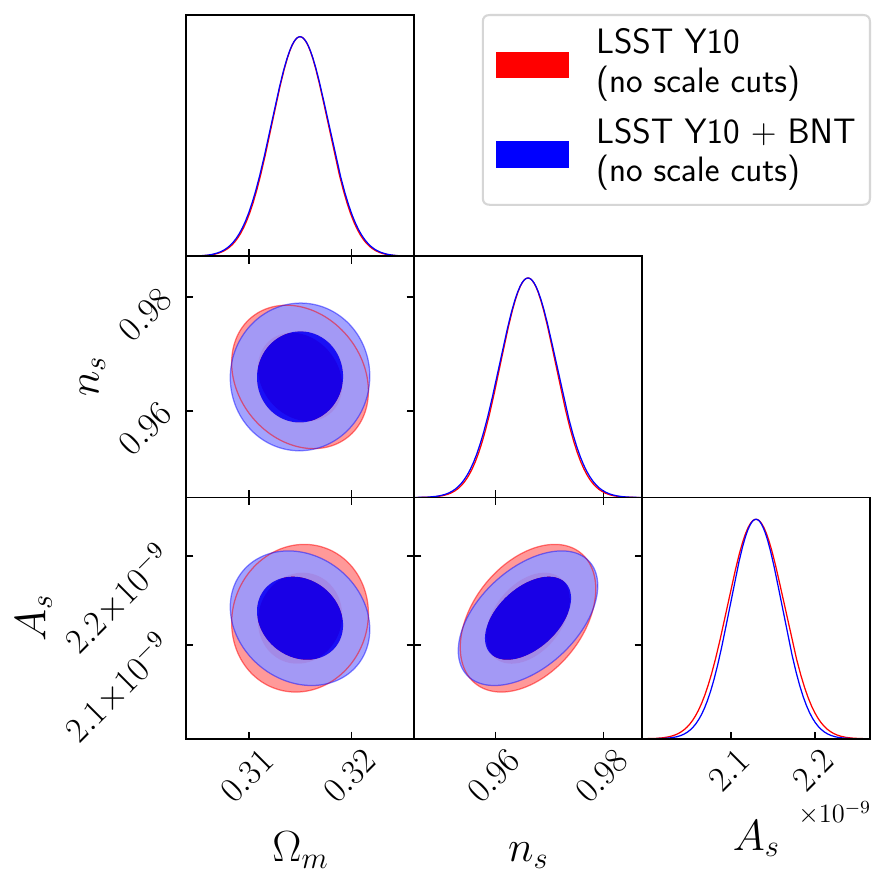}
    \caption{ We show that constraining power is unchanged irrespective of whether the BNT transform is applied in the case where no scale cuts are applied to the data vectors. This is to be expected since the BNT is essentially a linear and invertible matrix transformation. This result is to be viewed as a sanity check of our pipeline and of the likelihood sampling method of estimating the covariance matrix \cite{Taylor_2018}.}
    \label{fig:sanity_check}
\end{figure}

As an additional sanity check on our results, we check that that the covariance matrix that we compute is consistent with the data our pipeline produces independently. We do this by computing many mock samples of the $k-$cut $C(\ell)$ using Cholesky decomposition of the covariance matrix. We then average over all these realisations and then compare the distribution of these $C(\ell)$ with the BNT transformed ones computed by our Cosmosis pipeline. We show in the bottom panel of fig.~\ref{fig:sanity_check} the results of our $\Lambda$CDM LSST Y10 Fisher forecast with no scale-cuts employed, where the BNT transformation should have no effect on the constraining power. This is borne out in the Fisher contours, as expected.  

\subsection{Fisher matrix forecasts with \texttt{Cosmosis}} \label{sec:cosmosis}

We perform Fisher forecasts on the binned parameterisation  with a simulated data vector, assuming an LSST Y10-like survey. For the source and lens redshift distributions, we use the well-known Smail formula 
\begin{equation}\label{eq:smail}
 n(z) \propto \left(\frac{z}{z_0}\right)^{2}\exp\left[-\left(\frac{z}{z_0}\right)^{\beta}\right]\,.
\end{equation}
For LSST Y10, we use $\beta = \{0.68, 0.90\}$, $z_0 = \{0.11, 0.28\}$ and $n_{\rm gal} = \{27.0, 48\}\,{\rm arcmin}^{-2}$. We smooth the $n(z)$ kernels with a Gaussian that quantifies photo-$z$ uncertainty, which we set $\sigma_{\rm z} = 0.05 (1+z)$. We assume a sky fraction $f_{\rm sky} = 0.35$. We then use the Limber approximation to compute the projected signal for the $3\times 2$pt observables given by 
\begin{equation}\label{eq:Cl}
    C_{ij}^{XY}(\ell)=c\int_{z_{\rm min}}^{z_{\rm max}}\text{d}z\frac{W_i^X(z)W_j^Y(z)}{H(z)r^2(z)}P(k_{\ell},z)\, ,
\end{equation}
where $k_{\ell}=(\ell+1/2)/r(z)$, $r(z)$ represents the comoving distance as a function of the redshift, and $P(k_{\ell},z)$ stands for the non-linear matter power spectrum evaluated at a wavenumber $k_{\ell}$ and redshift $z$. The kernels or window functions are given by
\begin{align}
    W_i^{\rm G}(k,z)=&b_i(k,z)\frac{n_i(z)}{\bar{n}}\frac{H(z)}{c}\, ,\\
     W_i^{\rm L}(k,z) =& \frac{3}{2}\Omega_{\rm m} \frac{H_0^2}{c^2}(1+z)\,r(z)\,\Sigma(z)
\int_z^{z_{\rm max}}{\text{d} z'\frac{n_i(z')}{\bar{n}_i}\frac{r(z'-z)}{r(z')}}\nonumber\\
  &+W^{\rm IA}_i(k,z)\, ,
\end{align}
where the `G' and `L' labels signify clustering and lensing, respectively and $\Sigma = \frac{1}{2}\mu \left[1+\eta \right]$. The ratio $n_i(z)/\bar{n}$ represents the normalised galaxy distribution as a function of redshift, while $b_i(k,z)$ is the galaxy bias in the $i$-th tomographic bin. Note that in our pipeline, we just consider vary the linear bias parameter in each bin assuming no additional scale/redshift dependence. The effects of modified gravity will be encapsulated in $\Sigma(z)$ for the lensing potential and the matter power spectrum $P(k_{\ell},z)$. The intrinsic alignment contribution to the weak lensing kernel enters through the $W_i^{\rm IA}(k,z)$ term. We consider the non-linear alignment model with the single free parameter being the amplitude of intrinsic alignments, $A_{\rm IA}$ \cite{Kirk_2012_IA, Bridle_2007}. We remark that while this assumption may be overly simplistic, it is important to note that no detailed study of intrinsic alignment including higher order tidal terms in the context of modified gravity. Therefore, we leave a more detailed treatment of intrinsic alignment (and for the same reasons, more sophisticated treatment of the galaxy bias) to future work. In table \ref{tab:params}, we present the set of parameters that we vary in our Fisher forecasts and their fiducial values. 

\begin{table}
\centering
\begin{tabular}{|c|c|}
\hline
Parameters $\Theta$ & Fiducial Value  \\ 
 \hline
 $\Omega_{\rm m}$ & 0.315  \\
 \hline 
 $\Omega_{\rm b}$ & 0.049   \\
 \hline
 $n_{\rm s}$ & 0.965 \\
 \hline
 $A_{\rm s}$ & $2.13\times 10^{-10}$  \\
 \hline
 $h$ & 0.674\\
 \hline
 $\log T_{\rm AGN}$ & 7.8   \\
 \hline
 NLA $A_{\rm IA}$ & 1.0 \\
 \hline
 linear bias $b_k$ & 2.0 \\
 \hline
 $\mu_i$  & 1.0 \\
 \hline
 $\eta_i$ & 1.0 \\
 \hline
 photo-z lens errors $\vartheta_k$ & 0. \\
 \hline
 photo-z source errors $\varphi_i$ &  0. \\
 \hline
\end{tabular}
\caption{Fiducial parameter values for Fisher forecasts, assuming an LSST Y10-like survey. Note that the index $k$ is a label for the tomographic bins for the lens galaxies of which there are 10, while the index labels the source galaxy bins, of which there are 5. }
\label{tab:params}
\end{table}

The Fisher information matrix is formally defined to be the expectation value of the second derivatives of the log-likelihood function relative to the parameters and is typically interpreted as the curvature of the likelihood function around the maximum. For a Gaussian likelihood and a parameter-independent analytical covariance matrix $C_{ij}$, we have 
\begin{equation}
    F_{\alpha\beta} = \frac{\partial d_i}{\partial \Theta_{\alpha}}C^{-1}_{ij}\frac{\partial d_j}{\partial \Theta_\beta}\, .
\end{equation}
\texttt{Cosmosis} has been extensively used to compute Fisher matrices \cite{Harrison_2016, ref:EuclidWL}. We ensure the numerical stability of the derivatives by checking that the Fisher matrix elements are robust to choice of the step-size and method of derivative calculation. Once the Fisher matrix has been computed, one can obtain the one-sigma error on each parameter is given by the diagonal elements of the inverse Fisher matrix $F^{-1}_{ij}$. Of course, what we are interested in is the marginalised 1-sigma errors on the cosmological and modified gravity parameters. We compute the marginalised constraints by simply removing the rows and columns corresponding to the nuisance parameters from the inverse Fisher matrix. We define the Fisher figure-of-merit as the $-M/2$th root of the determinant of the Fisher matrix where $M$ is the subset of parameters obtained \textbf{after marginalisation}, i.e., 
\begin{equation}\label{eq:FOM}
    \text{Fisher FOM} = -\log_{10}(|F_{ij}|^{-0.5/M}) \, .
\end{equation}
The standard choice for the Fisher FOM is the square root of the determinant. However, we choose to take $M/2$ root in order to punish over-fitting/artificial increase in the determinant from an increase in parameter volume without a substantial gain in constraining power. We will further discuss and contextualise what an increase in the FOM translates to in terms of parameter constraints in sec.~\ref{subsec:quant_constraints}. 

We implement \texttt{ReACT} and \texttt{isitGR} into independent \texttt{cosmosis} modules that we use to compute the background cosmology and the non-linear $P(k)$. We then compute the boost factor due to baryonic feedback via the \texttt{HMCode} parameter $\log_{10}T_{\rm AGN}$ for the same set of cosmological parameters (independent of the modified gravity parameters). This allows us to compute the power spectrum suppression assuming that there is no coupling between the baryonic sector and the modified gravity phenomenology. Such a decoupling has been seen in simulations of $w$CDM cosmologies \cite{Mead_2015}. We will return to the this point later. We compute the $3\times 2$pt observables assuming the Limber approximation, the NLA model for intrinsic alignment, and bin-wise linear bias (see sec.~\ref{sec:cosmosis} for a more detailed analysis and justification of the modelling choices). We assume an analytical covariance and a Gaussian likelihood when computing the Fisher matrix elements.

\section{Results}\label{sec3}

\begin{figure}
    \centering
    \includegraphics[width=0.45\linewidth]{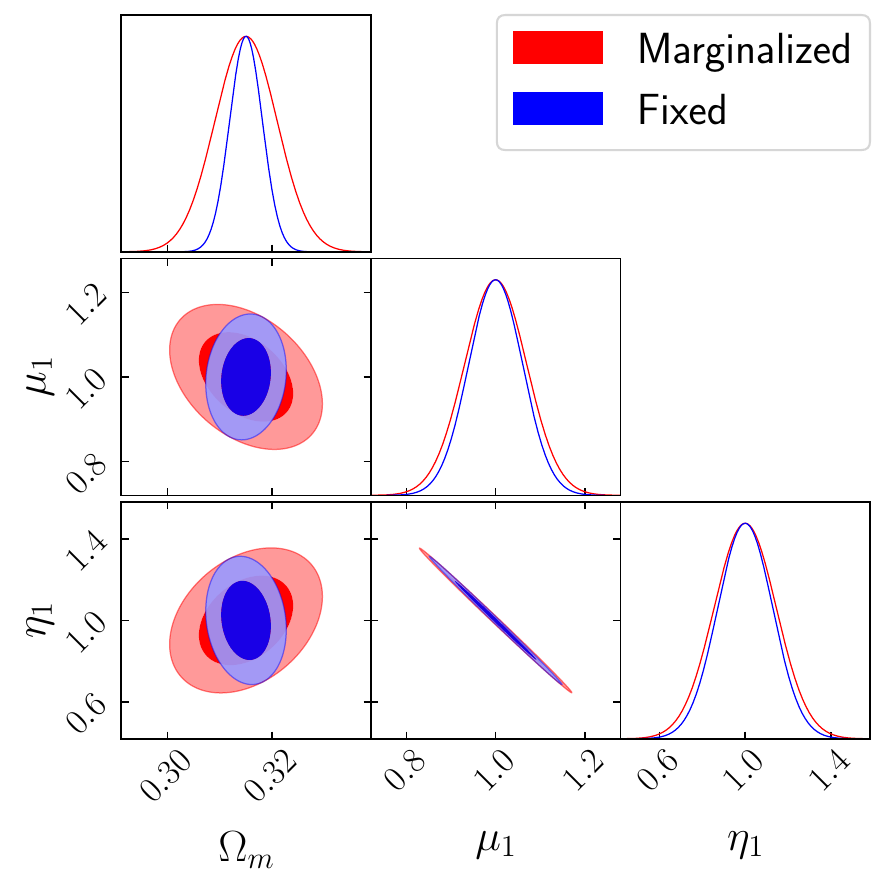}
    \includegraphics[width=0.45\linewidth]{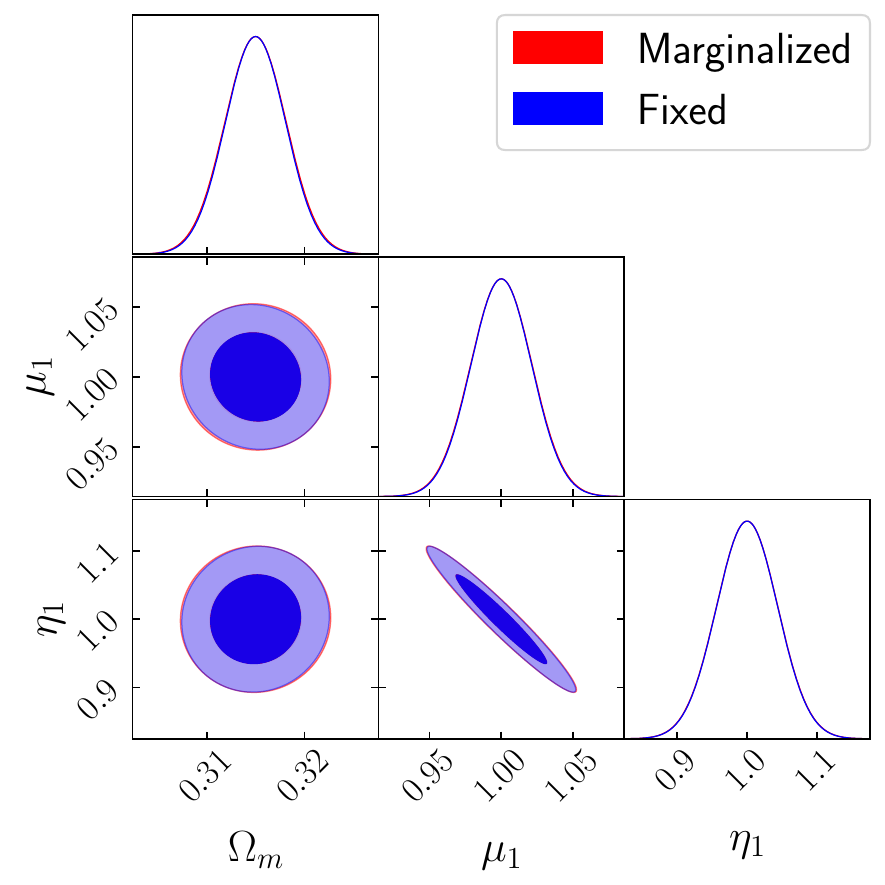}
    \caption{We show the Fisher contours for the subset $\{\Omega_{\rm m}, \mu_1, \eta_1\}$ when we marginalize over the baryonic feedback parameter compared to when we fix it to its fiducial value with the scale cut that correspond to $k_{\rm cut} = 0.5\,h\,{\rm Mpc}^{-1}$. In the left panel, we see that the $\Omega_{\rm m}-T_{\rm AGN}$ degeneracy is much larger compared to the degeneracy between $T_{\rm AGN}$ and the MG parameters. In the right panel, we repeat the same procedure but with the BNT transformed data vector and find a negligible effect. This is due to the fact that BNT transformation allows greater control over the scales that enter into the $C(\ell)$ computation (see fig.~\ref{fig:lensing}).  }
    \label{fig:baryon_marg}
\end{figure}

We will begin our analysis by investigating the impact of the BNT transform and its utility in mitigating for small scale physics, in particular baryonic feedback, in our modified gravity case. We will then move on to the matter of determining how many bins one can reasonably constrain with an LSST Y10-like survey for the different modelling choices that we consider. Finally, we will present the marginalised Fisher constraints on the full set of cosmological and modified gravity parameters and show which modelling choices lead to the best constraints. We will then end the section by discussing the physical origins of the relevant degeneracies and comment on the impact choice of the modified gravity parameterisation.

\subsection{Impact of BNT transform: mitigating baryonic effects}\label{subsec:baryons}

One of the biggest challenges facing large scale structure in the coming decades is baryonic feedback. In order to quantify the effect of baryons on parameter constraints, we compute the Fisher matrix for two cases: one where we fix the AGN feedback parameter $T_{\rm AGN}$ to its fiducial value and one where we marginalize over it. We then repeat this procedure including the BNT transform in the modelling. 
By comparing the parameter constraints  with and without marginalisation, and with and without the BNT transformation, we can see the effect that uncertainty in the baryon modelling has on our analysis in each case. We set $k_{\rm cut} = 0.5\,h\,{\rm Mpc}^{-1}$ initially, and comment on relaxing this below. We present the results of this analysis in fig.~\ref{fig:baryon_marg}.

In our setup, i.e., where we model the feedback with a single parameter $T_{\rm AGN}$, there is a degeneracy between the baryon parameter and some of the MG parameters, but our results indicate that this degeneracy is much more significant for the $\Lambda$CDM parameters. The uncertainty on $\Omega_{\rm m}$ for example nearly doubles when one marginalises over the $T_{\rm AGN}$ compared to when it is fixed as shown in fig.~\ref{fig:baryon_marg} in the case where we do not make use of the BNT transform. It is worth pointing out that this may not be the case for a more flexible baryon model such as those motivated by a recent study of 'baryonification' using a combination of weak-lensing data in the dark energy survey (DES) and the kinematic Sunyaev-Zeldovic (kSZ) effect as seen by the Atacama Cosmology Telescope \cite{bigwood2024}. However, when we repeat this procedure for the BNT transformed data vector, we see that the effect of marginalising over baryons is negligible as shown in the right panel of fig.~\ref{fig:baryon_marg}. We note that this result is insensitive to the choice of whether or not we choose to include the concentration fitting function in our modelling. This is due to the greater control over the range of scales that enter into the weak-lensing kernel and indicates the power of the such nulling schemes.

We find that when we repeat this procedure for $k_{\rm cut} = 1\,h\,{\rm Mpc}^{-1}$, the error on the parameters increases to up to 100\%. In addition, the constraints in table \ref{tab:FisherResults} show that no parameters have worse constraints when the BNT transform is used. Combined, these results indicate that the BNT transform with $k_{\rm cut} - 0.5\,h\,{\rm Mpc}^{-1}$ should be used when doing a phenomenological modified gravity analysis with stage IV data, and that when doing so the analysis is not sensitive to baryonic feedback.

\subsection{Binning Schemes}\label{subsec:binning_schemes}

\begin{figure}
    \centering
    \includegraphics[width=0.7\linewidth]{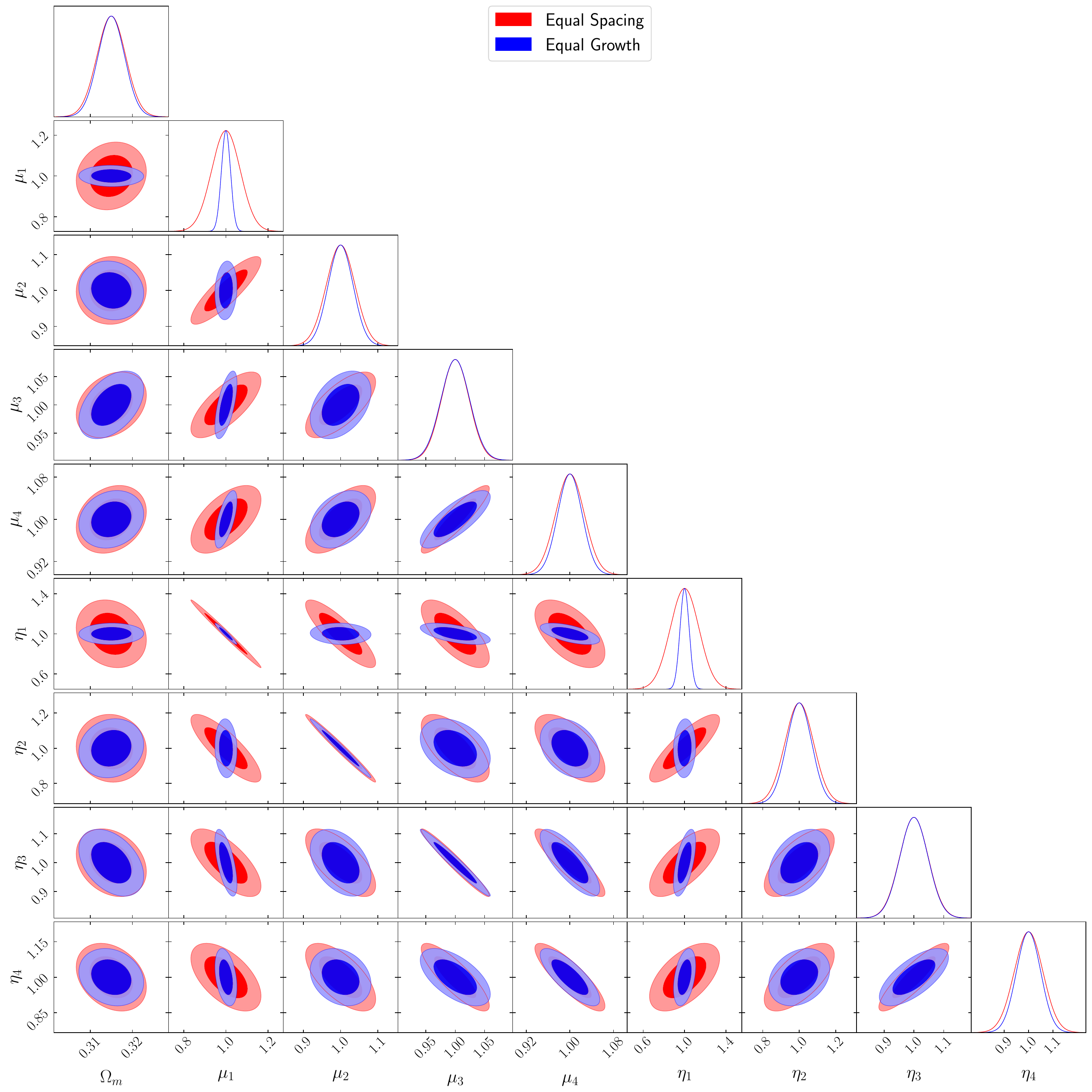}
    \caption{We present the modified gravity constraints for redshift bins of equal size in redshift (red) and equal increment incremental growth (blue). We see that equal growth bins are better constrained, as going to non-linear scales breaks degeneracies in this case that are still present in the case equal sized bins.  }
    \label{fig:equal_growth_vs_equal_size}
\end{figure}

Our initial study divides the redshift range in five bins in the range $0\leq z \leq 3$ (see table ~\ref{tab:bins}). Without loss of generality, we restrict ourselves here to the BNT transformed data vector with $k_{\rm cut} = 0.5\,h\,{\rm Mpc}^{-1}$ (we check and find that these results are robust to the choice of $k_{\rm cut}$ and the choices of including BNT/concentration fit in the modelling). We find that for these choices, we are able to constrain almost all modified gravity parameters to within 10\% (we will return to this point and quantitatively describe and discuss our results in the coming subsections), except for $\mu_5$ and $\eta_5$ ($2.15 \leq z \leq 3$). The error on this bin is degraded by over an order of magnitude relative to the other bins. This is due to the fact that the structure formation is linear in this redshift range on the scales that enter our Fisher forecast. As we are focusing on the impact of non-linear structure formation, we therefore fix $\mu = \eta = 1$ in this bin for the rest of this work. 

Since we now understand the redshift range in which the data is sensitive to modified gravity effects, we can now explore different binning schemes in an attempt to optimise constraining power. In \cite{Srinivasan2021, Srinivasan_2024}, we argued for redshift bins of equal incremental linear $\Lambda$CDM growth. The reason for this is that the bins naturally decrease size going from early to late times, which accounts for non-linearity at late times. Our expectation was that this would ensure that the constraining power on each bin would be roughly the same. In contrast to this choice, equal sized redshift bins have been studied in the literature \cite{ref:Casas2017}. We present the constraints on the modified gravity parameters for the two cases (again, we include the BNT transform in the modelling) in fig.~\ref{fig:equal_growth_vs_equal_size}. We find greater constraining power for the equal linear growth bins, and while the constraints are not completely uniform, the scatter in the $1-\sigma$ error on the parameters is significantly lower for the equal growth bins. 

\begin{figure}
    \centering
    \includegraphics[width = 0.45\linewidth]{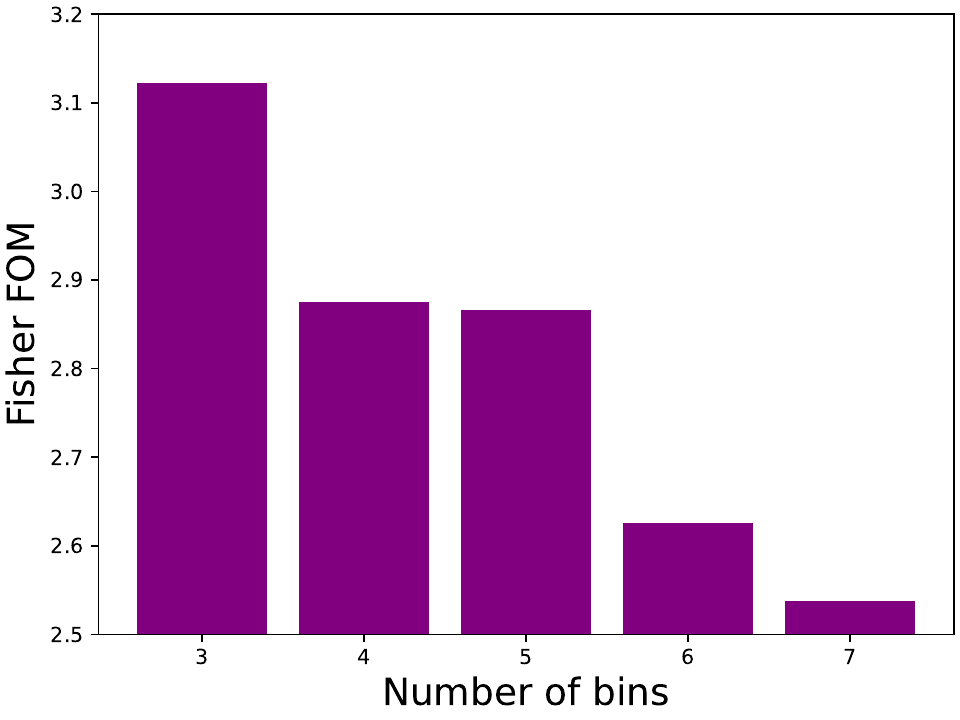}
    \includegraphics[width = 0.45\textwidth]{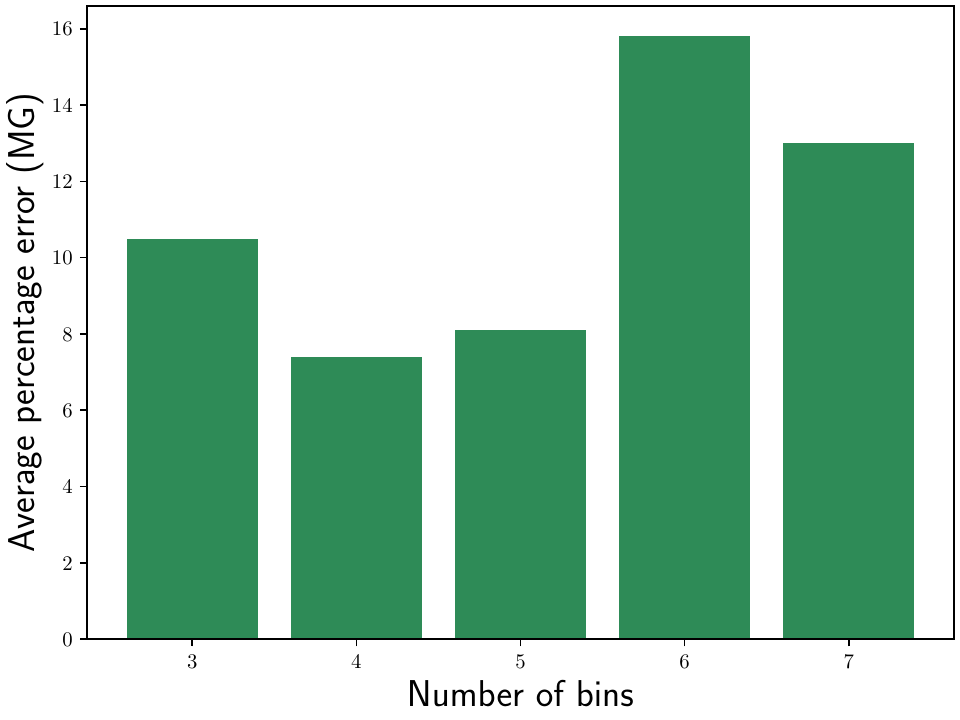}
    \caption{The Fisher FOM is shown as a function of MG redshift bins, i.e., for different choices of $D(z)$. We find that the FOM increases when going from 4 bins to 5 bins due to degeneracy breaking in the $\mu-\eta$ plane as shown in the right panel of fig.~\ref{fig:Cls_conc_degeneracy}. }
    \label{fig:FOM_bins}
\end{figure}

We now turn towards identifying the optimal number of bins for an LSST Y10-like survey. For this, we divide the redshift range between $0\leq z \leq 2.15$ into $N$ bins where  $N = \{3, 4, 5, 6, 7\}$. The bin-widths correspond to equal $D(z)$ (as computed in $\Lambda$CDM). Since the concentration fitting function has been validated for $0.8 \leq \mu \leq 1.2$ and $1.26 \leq D(z) \leq 1.45$, it cannot be included in the modelling for $N \geq 5$ bins since the corresponding $D(z)<1.26$. We use the Fisher FOM and the average constraining power on the MG parameters to statistically compare different binning schemes. We show the results of this procedure on fig.~\ref{fig:FOM_bins}. We find that we obtain the best modified gravity constraints for $N=4$ case, but the Fisher FOM for the $N=3$ bin case is larger due to the fact that the $\Lambda$CDM parameters are constrained better for this scheme. Since we find relatively good constraints on the cosmological parameters (see sec.~\ref{subsec:quant_constraints} for more details), we will adopt the $N=4$ case as our `best' case and will present our final results for this choice.\footnote{Note that we repeated this analysis for the 3 and 4 bin cases whilst including the effect of the concentration, and the qualitative results did not change.} We leave a detailed analysis of whether there is an even more optimal choice for the bin sizes and spacing to future work.

\subsection{Impact of modelling on constraining power} \label{subsec:quant_constraints}
We consider three modelling cases. The first being the case where the non-linear $P(k)$ is computed with \texttt{ReACT} and scale cuts are applied in the 'standard' way with $k_{\rm cut} = \{0.1, 0.5, 1.0\}\,h\,{\rm Mpc}^{-1}$ for clustering and galaxy-galaxy lensing. For the weak-lensing signal in a given tomographic bin, the data vector is cut at $\ell=\ell_{\rm cut}$. $\ell_{\rm cut}$ is defined to be the $\ell$ at which the convergence power spectrum computed from the cut matter power spectrum (i.e. with the corresponding $k_{\rm cut}$) disagrees at the level of $5\%$ with that computed from the full matter power spectrum.

In the second modelling case, we apply the BNT transform to the data vector before applying any scale cuts, followed by using the same three choices of $k_{\rm cut}$ for all tracers. Since the BNT transformation has re-weighted the kernels, this allows us to cut scales solely using the modelling of $P(k)$. Finally, in the last case, we include the concentration fitting function in the modelling of the non-linear $P(k)$ in addition to applying the BNT transformation.

We note that our most conservative choice of $k_{\rm cut} = 0.1\,h\,{\rm Mpc}^{-1}$ yields constraints that are essentially identical to a fully linear analysis for all modelling cases. This is expected as the matter power spectrum $P(k)$ essentially linear and therefore identical for all our modelling cases. 

The summary of our results is presented in table \ref{tab:FisherResults}, where we present the marginalised 1-$\sigma$ constraints on all the parameters of interest. As expected, the constraining power of the survey increases substantially as a function of $k_{\rm cut}$, particularly for $\mu_i$ that influence the non-linear shape of the power spectrum as shown in \cite{Srinivasan2021}. One can further improve constraining power on applying the BNT transform, which ensures that one has better control over the scales that enter into the data vector and the scales that are cut.

\begin{figure}
    \centering
    \includegraphics[width=0.7\linewidth]{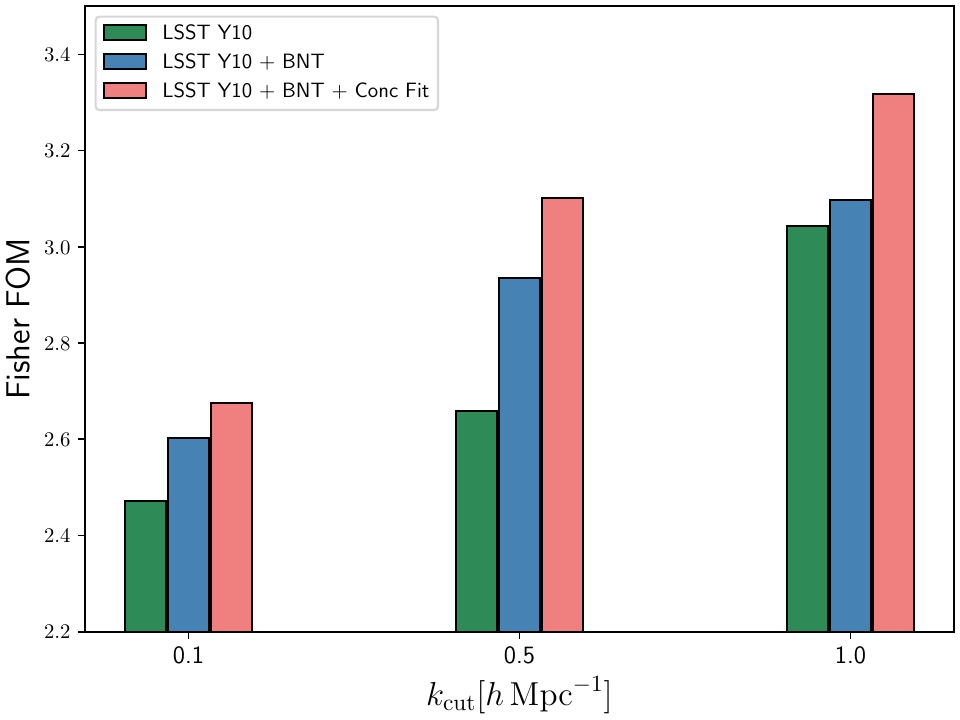}
    \caption{The Fisher figure-of-merit (see eq.~\eqref{eq:FOM} and the associated text for the explicit definition) for the different modelling cases considered in this work.  }
    \label{fig:FOM}
\end{figure}

However, our best constraints come from including the concentration-mass fit into the \texttt{ReACT} in addition to applying the BNT transform to the lensing $C(\ell)$. The main effect of including the fitting function into \texttt{ReACT} is that it adds additional shape information at low redshift, which greatly improves constraining power at $z\leq 1.0$. This results in significantly improved constraints in the first two modified gravity bins, and a much smaller improvement in the latter two modified gravity bins, as shown in table \ref{tab:FisherResults}. We show the quantitative improvement in the Fisher information when including the concentration fit and the BNT transform in fig.~\ref{fig:FOM}. 

For all of the cases, we gain nearly an order of magnitude in constraining power when going from an almost exclusively linear analysis with standard modelling to including all scales up to $k_{\rm cut} = 1.0\,h\,{\rm Mpc}^{-1}$, which corresponds to nearly a factor of $\sim 20$ in the average constraining power on the MG parameters. We found in \cite{Srinivasan_2024} the concentration fit improved the accuracy in reproducing $P(k)$ measured in the $N$-body simulations, especially for the low redshift bins. It is therefore no surprise that the constraining power associated to these bins improves from including it in the modelling.  

Moreover, as noted earlier, the (as of yet) unsolved problem of baryonic feedback prevents this case from becoming the automatic analysis choice going forward. We have shown however in fig.~\ref{fig:baryon_marg} that the baryonic effects can be mitigated in the case of $k_{\rm cut} = 0.5\,h\,{\rm Mpc}^{-1}$. For this scale cut with the BNT + concentration modelling, we are able to achieve an average constraint of $\sim 2.7\%$ on the MG parameters. We consider this to be the most realistic and best motivated choice to make.

\begin{figure}
    \centering
    \includegraphics[width=\linewidth]{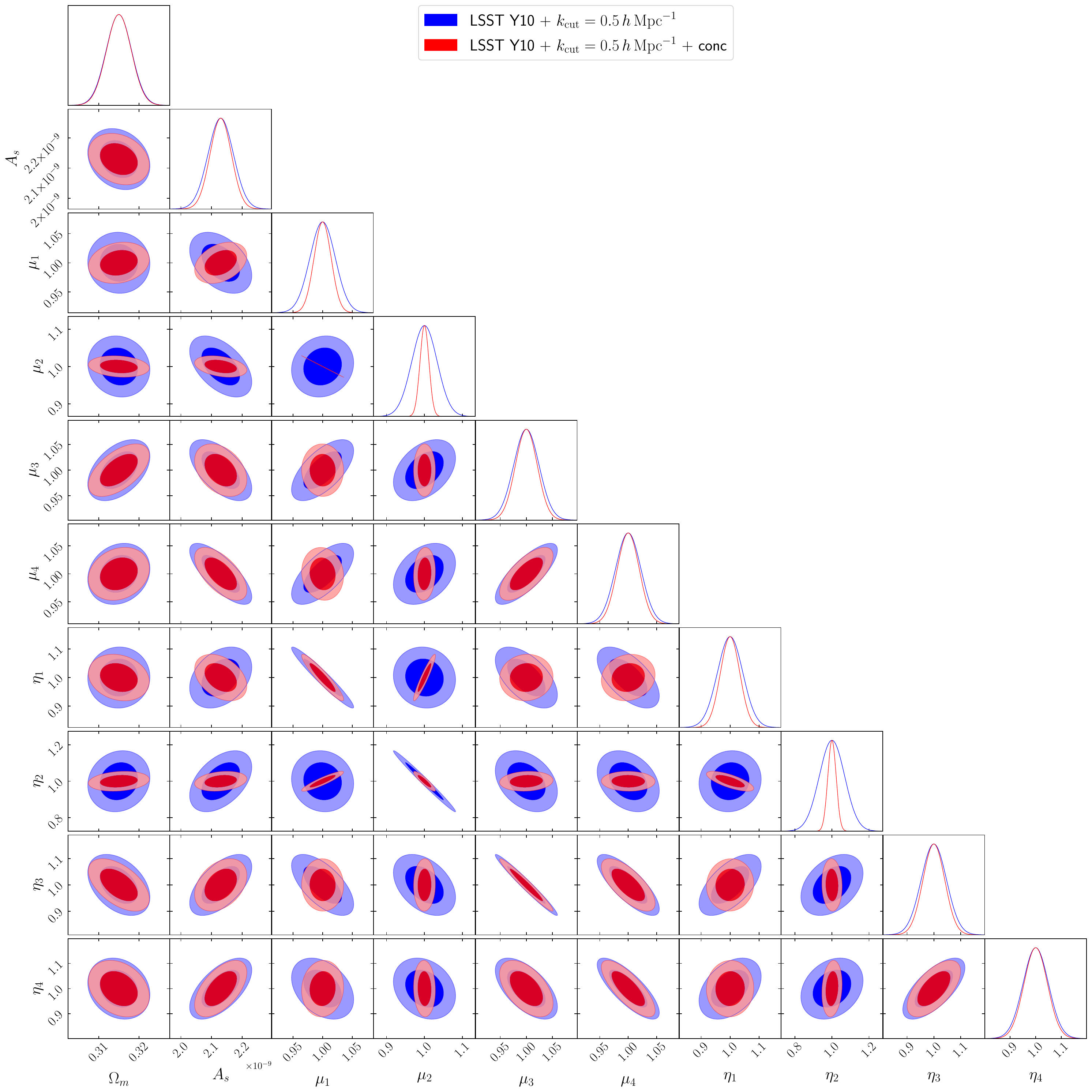}
    \caption{We present the $1-\sigma$ and $2-\sigma$ contours on the cosmological parameters and the modified gravity parameters for the middle and best different modelling cases described in the text (see table \ref{tab:FisherResults}). We see that including the additional information from the concentration-mass fit and the greater control of the scales that we include in the modelling allows us to not only have greater constraining power, but also breaks degeneracies between parameter combinations. }
    \label{fig:quasi_cut}
\end{figure}

\begin{table*}[ht]
\begin{center}
    \begin{tabular}{c|c|ccccc}
    \hline
    \hline
    Model &  $k_{\rm cut} [h\,{\rm Mpc}^{-1}]$ & $\Omega_{\rm m}$ & $\Omega_{\rm b}$ & $h$ & $n_{\rm s}$ & $A_{\rm s}$ \\
    \hline 
    \multirow{3}{*}{LSST Y10}		&	0.1  &	$3.0\%$	&	$3.9\%$	&	$20.0\%$	&	$3.6\%$	&	$12.5\%$		\\	
	  &	0.5	&	$1.8\%$	&	$2.8\%$	&	$14.7\%$	&	$1.8\%$	&	$2.2\%$	\\	
		&	1.0	&	$	1.2	\%$	&	$	2.3	\%$	&	$	6.1	\%$	&	$0.6\%$	&	$	2.0	\%$		\\ 
    \hline
     \multirow{3}{*}{LSST Y10 + BNT} & 0.1 &	$	2.6	\%$	&	$	3.0	\%$	&	$	15.6	\%$	&	$	2.0	\%$	&	$	6.2	\%$		\\	
      & 0.5 &	$	2.2	\%$	&	$	2.5	\%$	&	$	10.2	\%$	&	$	9.3	\%$	&	$	2.1	\%$		\\	
		&	1.0 &	$	0.6	\%$	&	$	2.0	\%$	&	$	3.1	\%$	&	$	0.5	\%$	&	$	1.5	\%$			\\ 
    \hline
    \multirow{3}{0.2\linewidth}{LSST Y10 + BNT + conc} & 0.1 &	$	2.2	\%$	&	$	1.2	\%$	&	$	12.7	\%$	&	$	1.9	\%$	&	$	5.4	\%$		\\	
      & $0.5$ &	$	1.0	\%$	&	$	2.2\%$	&	$	9.8\%$	&	$	0.7\%$	&	$	1.6\%$			\\ 
		&	$1.0$ &	$	0.5	\%$	&	$	1.6\%$	&	$3.0\%$	&	$0.3\%$	&	$	1.1\%$			\\ 
  \hline
  \hline
    \end{tabular}
    \begin{tabular}{c|c|cccc}
    Model &  $k_{\rm cut} [h\,{\rm Mpc}^{-1}]$ & $\mu_{1}$ & $\mu_{2}$ & $\mu_3$ & $\mu_{4}$  \\
    \hline 
    \multirow{3}{*}{LSST Y10}		&	0.1  &	$18.1\%$	&	$42.6\%$	&	$10.3\%$	&	$7.8\%$	\\	
	  &	0.5	&	$7.0\%$	&	$5.0\%$	&	$3.1\%$	&	$4.2\%$	\\	
		&	1.0	&	$	2.2	\%$	&	$	1.7	\%$	&	$	1.3	\%$	&	$	1.7	\%$	\\ 
    \hline
     \multirow{3}{*}{LSST Y10 + BNT} & 0.1 &	$	18.1	\%$	&	$	31.1	\%$	&	$	7.6	\%$	&	$	6.5	\%$	\\	
      & 0.5 &	$	5.2	\%$	&	$	3.3	\%$	&	$	2.5\%$	&	$	2.3	\%$	\\	
		&	1.0 &	$	1.5	\%$	&	$	1.4	\%$	&	$	1.0	\%$	&	$	1.4	\%$		\\ 
    \hline
    \multirow{3}{0.2\linewidth}{LSST Y10 + BNT + conc} & 0.1 &	$	17.5\%$	&	$	30.6\%$	&	$	7.1\%$	&	$	6.5	\%$	 \\	
      & $0.5$ &	$	1.4	\%$	&	$	1.2	\%$	&	$	2.1	\%$	&	$	1.9	\%$	\\	
		&	$1.0$ &	$	0.4	\%$	&	$	0.4	\%$	&	$	0.6	\%$	&	$1.3\%$	\\ 
  \hline
  \hline
    \end{tabular}
    \begin{tabular}{c|c|cccc}
    Model &  $k_{\rm cut} [h\,{\rm Mpc}^{-1}]$ & $\eta_{1}$ & $\eta_{2}$ & $\eta_3$ & $\eta_{4}$   \\
    \hline 
    \multirow{3}{*}{LSST Y10}		&	0.1  &	$36.7\%$	&	$85.7\%$	&	$29.8\%$	&	$18.3\%$ \\	
	  &	0.5	&	$14.5\%$	&	$10.1\%$	&	$6.1\%$	&	$10.6\%$	\\	
		&	1.0	&	$	4.1	\%$	&	$	2.5	\%$	&	$	2.0	\%$	&	$	4.0	\%$	\\ 	
    \hline
     \multirow{3}{*}{LSST Y10 + BNT} & 0.1 &	$	36.2	\%$	&	$	45.1	\%$	&	$	23.6	\%$	&	$	16.9	\%$		\\	
      & 0.5 &	$	10.5	\%$	&	$	6.3	\%$	&	$	5.2	\%$	&	$	6.3	\%$	\\	
		&	1.0 &	$	1.5	\%$	&	$	2.4	\%$	&	$	1.6	\%$	&	$	3.7	\%$	\\ 
    \hline
    \multirow{3}{0.2\linewidth}{LSST Y10 + BNT + conc} & 0.1 &	$	34.7\%$	&	$	44.7\%$	&	$	22.2	\%$	&	$	16.4	\%$	 \\	
      & $0.5$ &	$	3.3	\%$	&	$	2.2	\%$	&	$	4.1	\%$	&	$	4.6	\%$	\\	
		&	$1.0$ &	$	1.1\%$	&	$	1.0	\%$	&	$	1.5	\%$	&	$	3.5	\%$	\\ 
  \hline
  \hline
    \end{tabular}
\caption{Marginalised 68.3\% confidence level constraints on the free parameters. Values are reported in percentage of the parameters' fiducial values. Note that the relative increase in constraining power for the lower two redshift bins when including the BNT transform and the concentration-mass fit may be traced back to the shape of the fitting function for fixed $D(z)$ at low redshift, where the deviation from GR is the strongest (see fig.~\ref{fig:conc_fit} for the shape of the fit, fig.~\ref{fig:Cls_conc_degeneracy} for the effect on the lensing $C(\ell)$ and \cite{Srinivasan_2024} for more details on the fit). Note that the asterisk associated to the `standard' modelling case refers to the fact that the scale cuts we applied to weak-lensing data correspond to the $\ell_{\rm cut}$ value at which the convergence power spectrum computed with the corresponding $k_{\rm cut}$ disagrees at the level of $5\%$ with that computed from the full matter power spectrum, for a given tomographic bin. }
\label{tab:FisherResults}
    
\end{center}

\end{table*}
The results in table \ref{tab:FisherResults} are computed by marginalising over the nuisance parameters, which include the bias parameters $b_k$, the intrinsic alignment amplitude $A_{\rm IA}$, the photo-z uncertainties $\vartheta_k,\, \varphi_i$ and the AGN feedback parameter $T_{\rm AGN}$. To investigate the degeneracies between the parameters, we show in fig.~\ref{fig:quasi_cut} the 1 and 2$-\sigma$ contours for the parameter set $\{\Omega_{\rm m},\, A_{\rm s},\, \mu_{\rm 1}, \, \mu_{\rm 2},\, \mu_{\rm 3},\, \mu_{\rm 4},\, \eta_{\rm 1},\, \eta_{\rm 2},\, \eta_{\rm 3},\, \eta_{\rm 4}\}$ for the choice of $k_{\rm cut} = 0.5\,h\,{\rm Mpc}^{-1}$, i.e., the case where we push the modelling to mildly non-linear scales while being insensitive to baryons. We see that the most striking degeneracy in this parameter space is between $\mu_i$ and $\eta_i$ for the same redshift bin $i$. In this case, there is no change in the degeneracy direction even when including the concentration fit in the modelling. We will in what follows, justify the increase in constraining power for the BNT + concentration case as well as explicitly show the reason for the aforementioned degeneracy.

\subsection{Influence of degeneracies on modelling choices and binning strategies}\label{subsec:degeneracies}

\begin{figure}
    \centering
    \includegraphics[width = 0.45\textwidth]{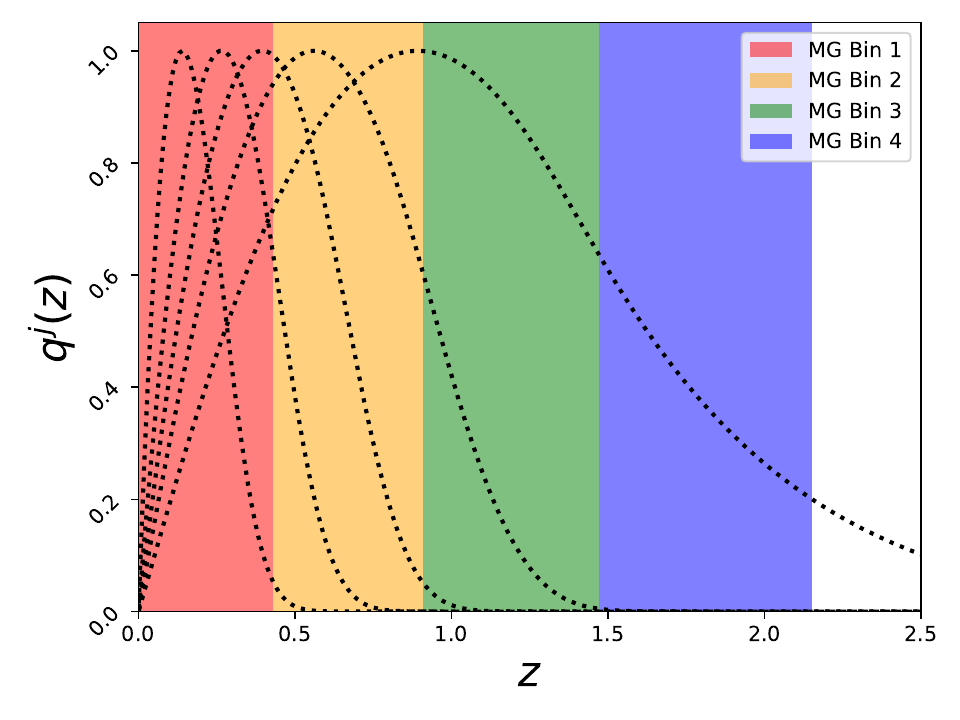}
    \includegraphics[width=0.45\linewidth]{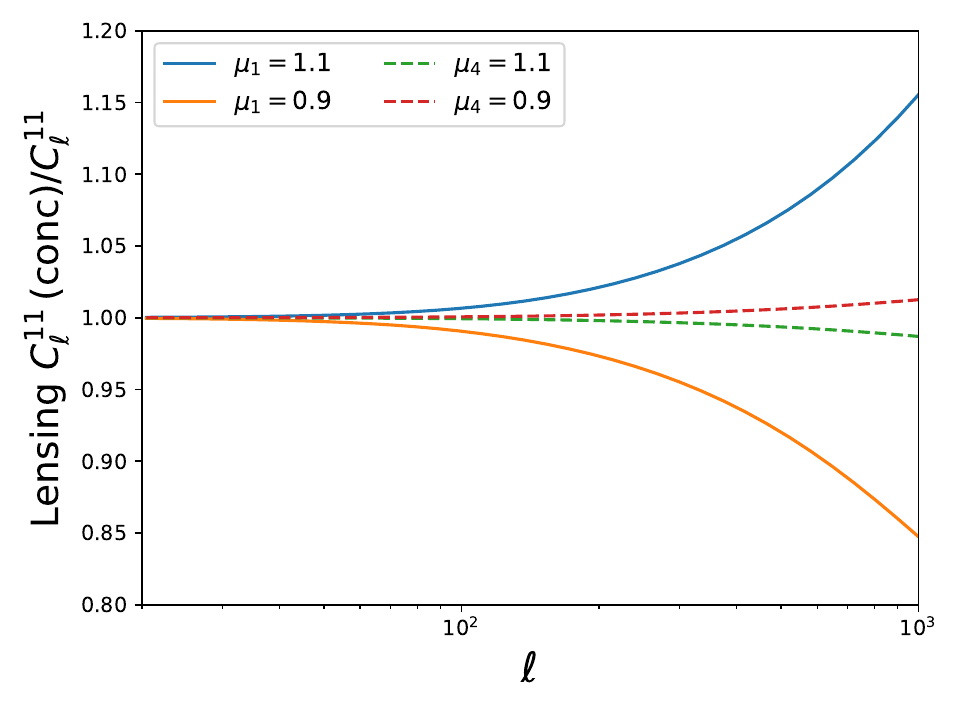}
    \caption{ In the left panel, we show the LSST Y10 source kernels of all five tomographic bins (black dotted). In the background, we colour code the redshift range the MG parameter bins (see table \ref{tab:bins} and associated text for details on why this binning choice was made). The constraining power on the modified gravity parameters is proportional to the information content that is available to the five tomographic bins weighted by the combined lensing kernel. Therefore, switching on MG at $z\sim1$ ensures maximum sensitivity to the lensing kernel, which is why $\mu_3$ and $\eta_3$ are our best constrained parameters in the absence of the concentration fitting function. In the right panel, we show the ratio of the first tomographic bin of lensing power spectrum when the concentration fit is included in the modelling relative to when it is not for $\mu_1$ and $\mu_4$ (red and blue in left panel). We see that when $\mu$ is modified at late times, the effect of the concentration fit is much larger, resulting in the increase in constraining power on these bins exclusively as seen in table \ref{tab:FisherResults} and fig.~\ref{fig:quasi_cut}. }
    \label{fig:Cls_conc_degeneracy}
\end{figure}

In order to understand the physics that drives the results in table \ref{tab:FisherResults} and make informed choices on the binning strategies for stage IV surveys, it is instructive to compare the data vector for different representative scenarios for $\mu$ and $\eta$. We will first evaluate what determines the constraining power on the MG parameters. We showed in fig.~\ref{tab:FisherResults} that $\mu_3$ and $\eta_3$ are the best constrained parameters in first and second modelling cases that do not include the concentration fitting function. This may be understood when you compare the span of the lensing kernels to the the redshift range of each bin. We show this in the left panel of fig.~\ref{fig:Cls_conc_degeneracy}. Essentially, it is bin 3 (in green) that is active (switched on) around where the lensing kernel associated the highest tomographic bin peaks. This ensures that every subsequent tomographic bin plays an important role in constraining $\mu_3$ and $\eta_3$ (and to a lesser extent, $\mu_4$ \& $\eta_4)$). On the other hand, the lower redshift bins are switched on later, and thus there is less information available, leading to weaker constraints. We note that this argument also makes clear how scale cuts we are using prevents the 5th modified gravity bin (see sec.~\ref{subsec:binning_schemes}) that we chose to not include in the modelling from being well-constrained.    

However, the inclusion of the concentration fit changes this bin hierarchy such that $\mu_1$ and $\eta_1$ are the best constrained parameters. In the right panel, we compare the effect of the concentration fit on the lensing $C(\ell)$ between the lowest redshift bin and the highest bin (see table \ref{tab:bins}). The additional constraining power on the lower redshift bin is due to this additional information in the data vector, which can be traced back to results from $N$-body simulations. These indicate that modifying $\mu$ at low redshift strongly influences non-linear structure formation (which is intrinsically more complex at late times). Therefore, $\mu_1$ and $\eta_1$ become the strongest constrained parameters when the concentration fit is included.

As we mentioned before, irrespective of the BNT/concentration modelling, the most prominent degeneracy in fig.~\ref{fig:quasi_cut} occurs between the MG parameters in the same redshift bin $\mu_i$ and $\eta_i$ (along the diagonal). This is because the lensing observable is shifted up identically on linear scales when $\eta-1 = 2(\mu-1)$, which can be reconciled to the weighting factor $\Sigma = \mu(1+\eta)$ in the lensing $C(\ell)$ (see eq.~\eqref{eq:Cl}). This degeneracy therefore is stronger at low redshift, since the lensing signal dominates the $3\times2pt$ data vector at low redshift (the clustering kernel is effectively zero at $z<0.2$). As we go from low to high redshift, although the lensing itself increases in constraining power (since this is where the kernels for both lensing and clustering peak), the relative importance of the clustering increases as well, resulting in the degeneracy being weaker. This can be seen in fig.~\ref{fig:quasi_cut} where the 2D $\mu_i$-$\eta_i$ ellipse is narrower and more elongated for $i=1,2$ than for $i=3,4$.

Our concentration fit adds information at late times, that allows a stronger constraint on $\mu_1$ and $\mu_2$ which shrinks the degeneracy, but doesn't eliminate it completely. This shrinking occurs because as one goes from linear to non-linear scales, degeneracies are broken due to the non-linear shape of $P(k)$ being dependent purely on $\mu$. However, this is a subtle effect that depends on the integrated growth associated with a given redshift bin. Of course, it is also important to bear in mind that the `extent' to which this degeneracy is broken depends on the non-linear shape of $P(k)$ which was shown to be a function of $\left(\mu, D(z)\right)$ in \cite{Srinivasan_2024, Srinivasan2021} (see also fig.~\ref{fig:enter-label} attached text in appendix \ref{app:degeneracies}). Going forward, the main challenge is to address the diagonal degeneracies between $\mu_i$ and $\eta_i$, which essentially involves incorporating additional information on large scale structure at $z\leq 0.5$.

In addition to this diagonal degeneracy, we also note that the only off-diagonal degeneracy that persists for all our modelling cases is $\mu_3-\mu_4$. In general, as can be seen in fig.~\ref{fig:quasi_cut},  applying the concentration on top of BNT only affects the degeneracies associated with either of the first two modified gravity bins ($\mu_{1-2}$ and $\eta_{1-2}$). For our preferred case of BNT+concentration+$k_{\rm{cut}}=0.5\,h\,$Mpc$^{-1}$, there are small degeneracies between the modified gravity parameters and the $\Lambda$CDM parameters, which reduce the constraints on the $\Lambda$CDM parameters. The largest effect is for $\Omega_m$ for which the 1-$\sigma$ error increases from about 0.68\% without modified gravity to 1\% with modified gravity. We note that the $\Lambda$CDM parameters are all strongly constrained even when modified gravity is included. We also note that another advantage of the non-linear modelling and thus using a scale cut of $0.5h$Mpc$^{-1}$, is that the degeneracies with $\Lambda$CDM are much smaller than in linear theory.

We now turn towards the choice of parameterisation. If one were to re-parameterise from $(\mu, \eta) \rightarrow (\mu, \Sigma)$, the degeneracy in the lensing $Cls$ may be seen explicitly in the Fisher contours where $\mu_1, \mu_2$ and $\Sigma_1, \Sigma_2$ are almost perfectly degenerate, respectively, on linear scales. In this case, the constraints on $\mu_i$ and $\Sigma_i$ are comparable, as opposed to $\mu-\eta$ where $\mu$ is generally constrained better than $\eta$ (see appendix \ref{app:degeneracies} for more detail). Since the basis of the non-linear recipe in our work is in the validation of $P(k)$ with $N$-body simulations for $0.8 \leq \mu \leq 1.2$, and the resultant constraints on $\mu$ are stronger, we choose the $\mu-\eta$ parameterisation to present all of our results. This also ensures the future combined analyses with spectroscopic surveys yield optimal results. We discuss in appendix \ref{app:degeneracies} details on the Fisher constraints for the $\mu-\Sigma$ parameterisation.

\section{Conclusions: The Road to Stage IV Data}\label{sec4}

In this work, we have computed Fisher forecasts of modified gravity models parameterised by an effective gravitational parameter $\mu$ and slip parameter $\eta$ that are both binned in redshift. We assumed an LSST Y10-like survey with 5 source bins and 10 lens bins. The systematic effects we include in our analysis are linear bin-wise galaxy bias $b_k$, intrinsic alignment amplitude $A_{\rm IA}$ (assuming the NLA model) and baryonic feedback parameterised by $\log_{\rm 10} T_{\rm AGN}$. We have three modelling cases, the first being the `standard' case where we use the \texttt{ReACT} implementation developed in \cite{Srinivasan2021} to compute the matter power spectrum and $3\times2$pt data vector. In the second case, we also apply a nulling transformation to the data vector called the BNT transform \cite{Bernardeau_2014, Taylor_2018, Taylor_2021} in order to localise the lensing kernels, which allows us to employ scale cuts directly from our knowledge of the matter power spectrum (see \ref{sec2.2} for more details). In the final case, we make use of the concentration-mass fitting function developed in \cite{Srinivasan_2024} in the non-linear modelling, in addition to the BNT transform. We reiterate that the modelling of the non-linear matter power spectrum for modified gravity used in this work is fully validated against N-body simulations \cite{Srinivasan2021, Srinivasan_2024}.

Our results may be divided into three main branches. Firstly, (see sec.~\ref{subsec:baryons} for details) we investigated the effectiveness of the BNT transform to mitigate the impact of baryonic feedback. We find that marginalising over $T_{\rm AGN}$ results has a big impact on the $\Lambda$CDM parameters ($>50\%$ for $\Omega_{\rm m}$) and a smaller $\sim 10\%$ increase in the error on the MG parameters assuming a $k_{\rm cut} = 0.5\,h\,{\rm Mpc}^{-1}$ for the clustering and galaxy-galaxy lensing signals and a corresponding $\ell_{\rm max}$ for the lensing signal. We show in the right panel of fig.~\ref{fig:baryon_marg} that this is mitigated for all parameters by the use of the BNT transform due to the control over the scales that enter the lensing kernel. This shows the strength and importance of such nulling schemes in this modified gravity context, particularly due to the large uncertainty on the specific model of baryonic feedback.   

Secondly, we studied different binning schemes for the modified gravity parameters with the aim of finding the optimal choice for a LSST Y10-like survey (see sec.~\ref{subsec:binning_schemes}). We find that constraining power on MG parameters sharply decreases at $z>2.1$. We therefore fixed the redshift range to be $0\leq z \leq 2.1$ and then divided this redshift range into $N=\{3, 4, 5,6, 7\}$ bins. While we found that Fisher FOM is maximal for $N=3$, we find that this is driven by an increase in constraining power on $\Lambda$CDM parameters, rather than the MG binned ones. We also showed that the bins are better and more equally constrained for bins of approximately equal linear growth. We find that the best average constraint of $\sim 7\%$ on the MG parameters is achieved for the $N=4$ bins case while still being able to constrain to achieve sub-10\% constraints on the $\Lambda$CDM parameters (see table \ref{tab:bins} to see what these bins are). 

We then presented our marginalised Fisher constraints on $\Lambda$CDM + MG parameters in table \ref{tab:FisherResults} for the three aforementioned modelling cases. We obtain the best constraints for the most sophisticated modelling (BNT + concentration) case and $k_{\rm cut} = 1.0\,h\,{\rm Mpc}^{-1}$, for which we achieve nearly an order of magnitude improvement in the Fisher FOM which (see eq.~\eqref{eq:FOM}) which translates to a factor $\sim 20$ improvement in the MG parameter constraints relative to the linear ($k_{\rm cut} = 0.1\,h\,{\rm Mpc}^{-1}$) case (see right panel of fig.~\ref{fig:Cls_conc_degeneracy} for more details on where this additional constraining power comes from). However, the caveat to this result is that we employ a simplistic one-parameter model for baryonic feedback in this work. We are still able to achieve sub $5\%$ precision in our constraints on $\mu$ and $\eta$ for a somewhat conservative choice of $k_{\rm cut}= 0.5 \,h\,{\rm Mpc}^{-1}$, which is the scale at which we showed the use of the BNT transform mitigates baryonic feedback effects. 

We show the 1$-$ and 2$-\sigma$ confidence intervals in fig.~\ref{fig:quasi_cut}, finding that in the absence of the concentration fitting, the best constrained MG parameters are $\mu_3$ and $\eta_3$. In  sec.~\ref{subsec:degeneracies} (see left panel of fig.~\ref{fig:Cls_conc_degeneracy}) we explain that this is due to the fact that this bin is switched on and off around where the lensing and clustering kernels peak. However, when we use the concentration fitting, we obtain that $\mu_1$ and $\eta_1$ are best constrained, which is due to the additional information from non-linear clustering entering the $Cl$ (see right panel of \ref{fig:Cls_conc_degeneracy}). We find that the dominant source of degeneracy is that between $\mu$ and $\eta$ in the same bin (along the diagonal in fig.~\ref{fig:quasi_cut}). Going to non-linear scales can break this degeneracy (see appendix \ref{app:degeneracies}) and varying $D(z)$, i.e., the bin-width and number of bins, can further help with decoupling $\mu$ and $\eta$. Finally, we comment on the differences between the $\mu-\eta$ and $\mu-\Sigma$ parameterisations. The $\mu-\Sigma$ parameterisation allows the lensing observable $\Sigma $ to be better constrained at the cost of a loss in constraining power on $\mu$. With possible future combined analyses with spectroscopic clustering surveys (which are sensitive only to $\mu$) in addition to the fact that our non-linear recipe is validated on $N$-body simulations with a redshift-binned $\mu$, we choose the $\mu-\eta$ parameterisation for our baseline analysis.

The pipeline we have used here could be deployed on real data, however there are some practical improvements that would make deploying the pipeline better and easier. One key improvement is the time required to compute the non-linear $P(k)$. Indeed, this has already been identified as a crucial barrier going forward with full Markov Chain Monte Carlo (MCMC) inference pipelines for Y5 stage III surveys taking several days to run on modern compute clusters, due to the number of times the Boltzmann solvers need to be called. One way to make progress is to create emulators for the matter power spectrum, of which many exist in $\Lambda$CDM \cite{Auld_2007, Fendt_2007, Lawrence_2017, Albers_2019, Mootovaloo_2022, Günther_2022, Euclid_emu2, Bacco_emu,  Bolliet_Cosmopower}. Extending this work to MG has been an active area of research \cite{Ramachandra_2021, Casares2023, fiorini2023}, although to date, the only MG models that have been considered have been $f(R)$/nDGP gravity. We remark that such an emulator based on the implementation of \texttt{ReACT} (note that such an emulator already exists for $f(R)$ \cite{Spurio_Mancini_2023}) in this work would be very useful to the community.  A second shortcoming to mention is that the detailed validation of \texttt{ReACT} against N-body simulations \cite{Srinivasan2021, Srinivasan_2024} has only been done for the case where one bin is switched on at a time. One can still do an MCMC analysis with a single bin whose position and width is varied, this would need to be extended for the more general case where multiple bins are switched on at once. Whilst one might expect the accuracy of the validated approach to broadly hold in the more general case, this should be checked in detail in order to be as rigorous as possible. We remark that approximate 2nd order Lagrangian Perturbation Theory (2LPT) solvers such as COLA \cite{ref:COLA} that have recently been used with some success in modified gravity could be very useful for such validation procedures.

The next decade will be a very exciting period for large scale structure cosmology, as data from stage IV surveys will become available. The work presented in this paper shows the value of applying this pipeline to data in order to comprehensively test $\Lambda$CDM. It also shows the value of carrying out the work to improve the pipeline identified above, and the range of parameters and analysis choices that should be focused on when creating emulators and when performing the data analysis. In particular, we recommend that the default analysis choices for phenomenological modified gravity for stage IV surveys should be $k_{\rm{cut}}=0.5h$Mpc$^{-1}$, 4 modified gravity bins of equal growth in the redshift range $0-2.1$, and the use of both BNT and the full concentration fitting formula adjustment to ReACT \cite{Srinivasan_2024}. Following this work, we are now close to model independent modified gravity constraints from non-linear scales with real data. 

\bibliography{ref.bib}
\bibliographystyle{apsrev4-1}

\appendix
 
\section{Origin of degeneracies and impact of parameter constraints}
\label{app:degeneracies}
\subsection{Data vectors}

\begin{figure}
    \centering
    \includegraphics[width=0.32\linewidth]{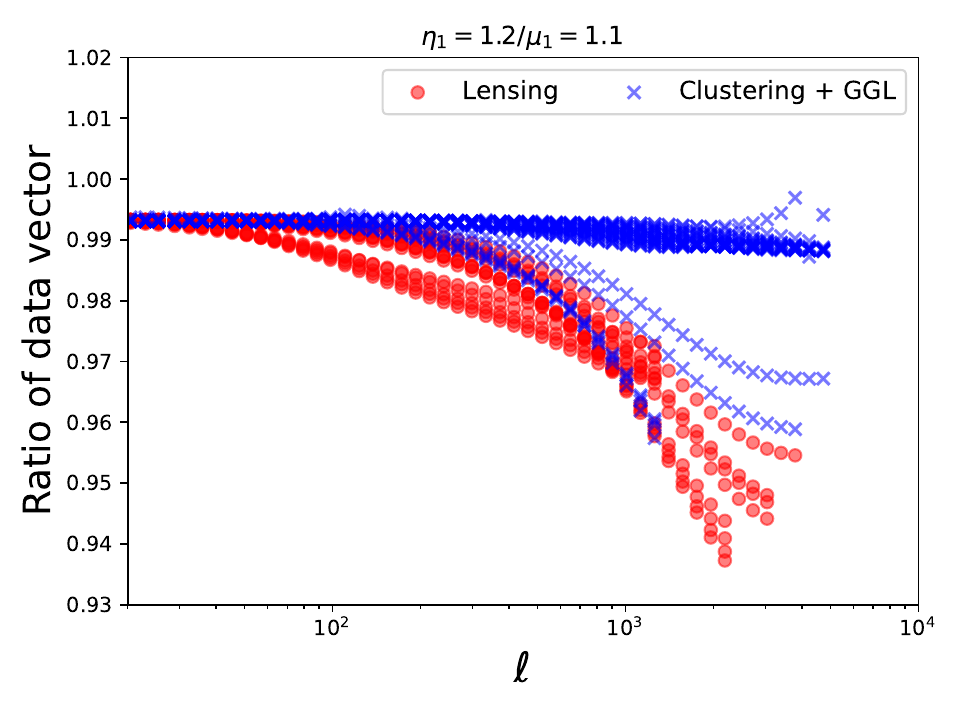}
    \includegraphics[width=0.32\linewidth]{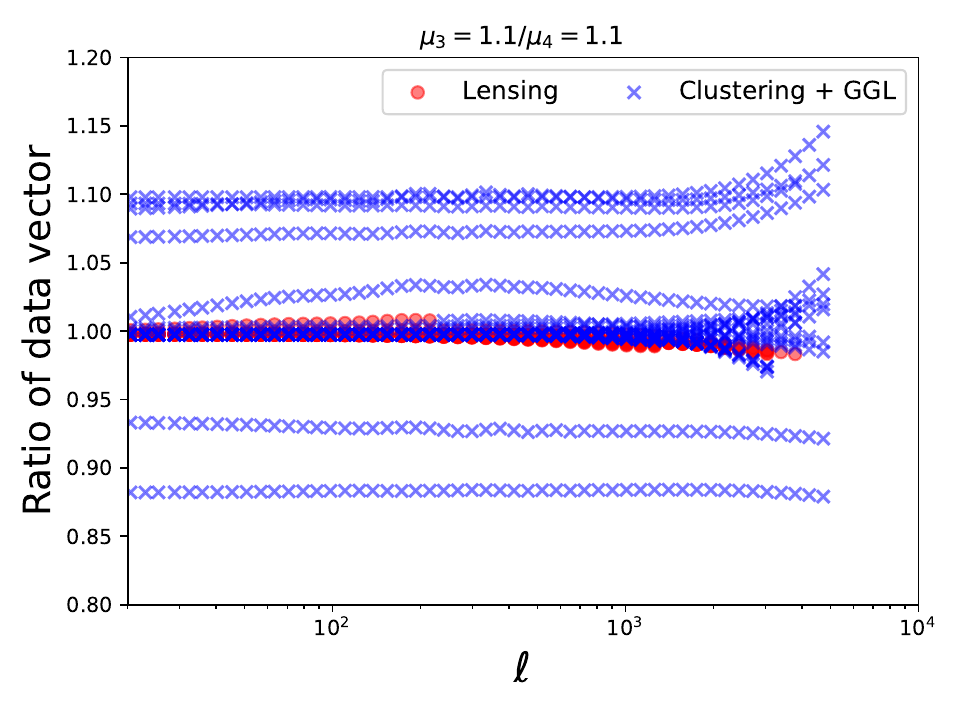}
    \includegraphics[width=0.32\linewidth]{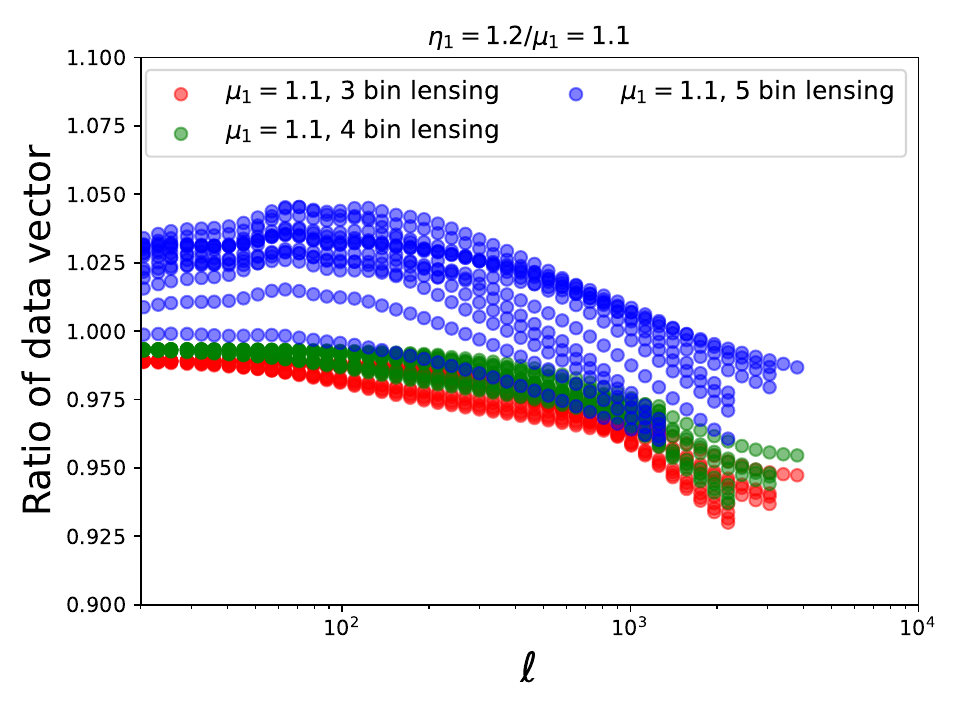}
    \caption{We show the ratio of degenerate data vectors, split into weak-lensing (red) and clustering + GGL (blue). We see that the lower end of our redshift range ($\mu_1 - \eta_1$), the lensing signal is dominant source of degeneracy breaking, going from linear to non-linear scales. However, going to higher redshift shows that clustering and GGL largely breaking degeneracies (in this case between $\mu_3 - \mu_4$). In the last panel, we show the ratio of the lensing signals for a varying MG bin number, showing that different bin widths can help to break MG parameter degeneracies.   }
    \label{fig:degeneracies}
\end{figure}

In this appendix, we explore the impact of different modelling choices on the $3\times2$pt data vector which leads to the results we showed in sec.~\ref{sec3} of the main text. We begin by computing the ratio of the data vector for the different probes we consider. We note that at the extreme low end of the redshift range we are interested in, the lensing signal dominates. In addition to this, it is well known that lensing is more sensitive to non-linear structure formation. We also note that the lensing kernels have a greater span in redshift ($0\leq z \leq 2.5$), while clustering plays an important role at $0.2 \leq z \leq 1.5$. Therefore, we would expect signal from clustering and the galaxy-galaxy-lensing (GGL) to play an important role in constraining power and breaking degeneracies at the intermediate redshift range. 

We show this in fig.~\ref{fig:degeneracies}, where in the left panel we compute the ratio of the data vector when $\mu_1 = 1.1$ relative to the case where $\eta_1 = 1.2$, i.e., along the main degeneracy direction (see fig.~\ref{fig:quasi_cut} and the discussion in sec.~\ref{subsec:degeneracies}), where we divide the data vector into the contribution from lensing and that from clustering + GGL. When we look at the higher redshift degeneracy between $\mu_3-\mu_4$ in the middle panel, we see that the clustering and GGL signals helps to break this degeneracy. Finally, we also show the $\mu_1-\eta_1$ degeneracy for different bin numbers (see sec.~\ref{subsec:binning_schemes} for details) in the right, where one clearly sees that varying the redshift range of bins that are degenerate with each other allows you to break such degeneracies. In the main text, our criteria for selecting the optimal binning scheme is to compare the average constraint on the MG parameters. We leave a more quantitative exploration of varying bin sizes to simultaneously optimise for constraining power and degeneracy minimisation to future work.

\subsection{Fisher contours}
\begin{figure}
    \centering
    \includegraphics[width=0.9\linewidth]{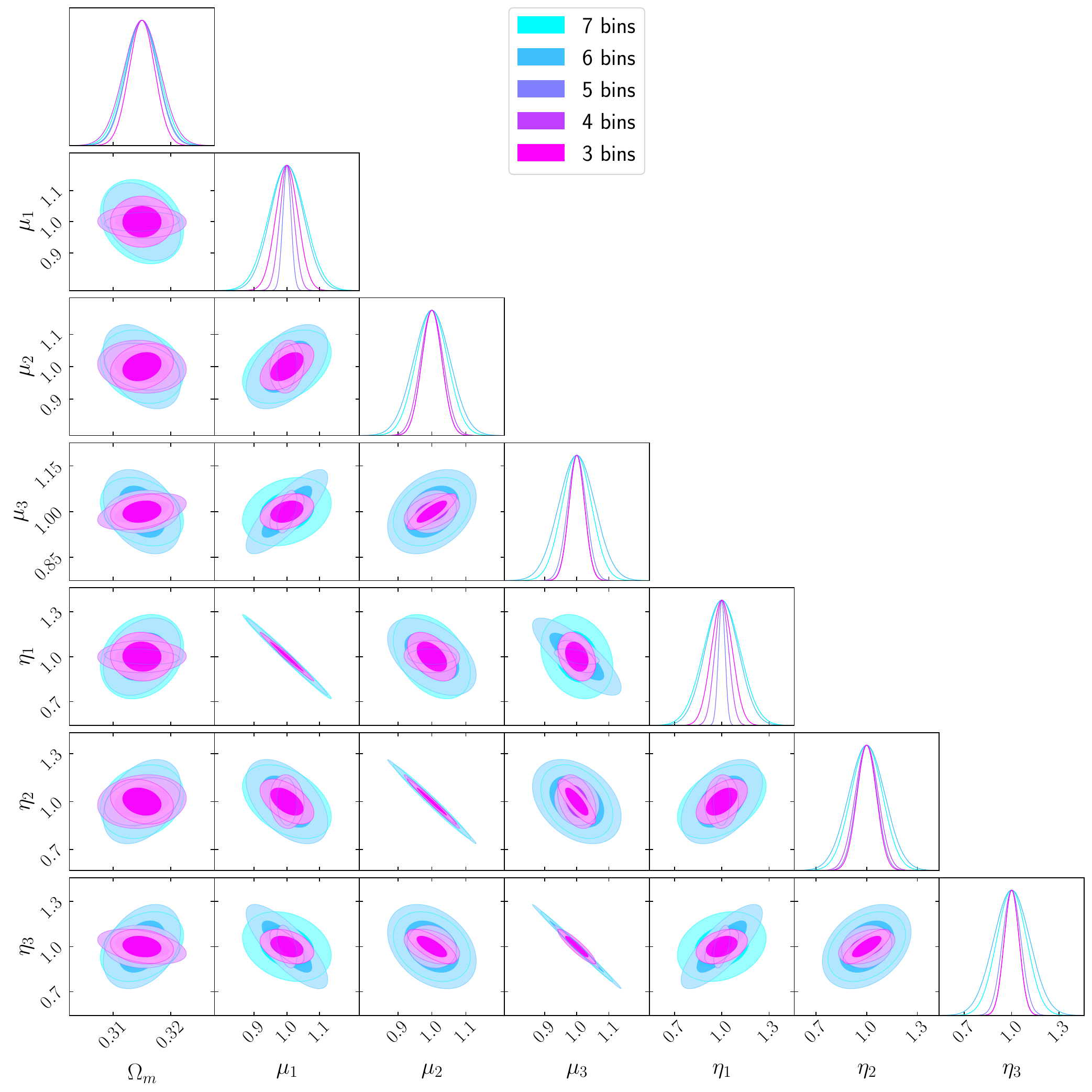}
    \caption{The fisher contours for the different binning schemes, i.e., the different values of of $D(z)$ associated to the modified gravity bins considered in this work (see sec.~\ref{subsec:binning_schemes} for more details).}
    \label{fig:enter-label}
\end{figure}

We now present the results of the Fisher matrix  forecasts that we condensed into the two panels in fig.~\ref{fig:FOM_bins}. Firstly, we present the Fisher contours as a function of bin number in fig.~\ref{fig:enter-label}. We find that the modified gravity parameters are best constrained for $N=4$ bins as shown in figs.~\ref{fig:Cls_conc_degeneracy} and \ref{fig:FOM_bins}, but this comes at a small cost to the constraining power on the cosmological parameters compared to using fewer modified gravity bins. It is also important to note that while the relative constraining power on each bin might improve on changing binning scheme, the \textit{average} constraining power tends to decrease as one adds more modified gravity bins at higher redshift, since this adds parameters that are poorly constrained by the data (relative to the previous set). This is quantitatively driven by the lack of change in the non-linear shape of the matter power spectrum when $\mu$ is modified at these redshifts (see \cite{Srinivasan_2024} for more details on the effect of $\mu(z)$ on $P(k)$). We also note that this analysis assumes that we only `switch on' one bin at a time (as is standard for a Fisher forecast with $\Lambda$CDM as the fiducial point), i.e. we do not consider the possibility that $\mu \neq 1$ in multiple bins. We leave this analysis to future work. 

\begin{figure}
    \centering
    \includegraphics[width=0.9\linewidth]{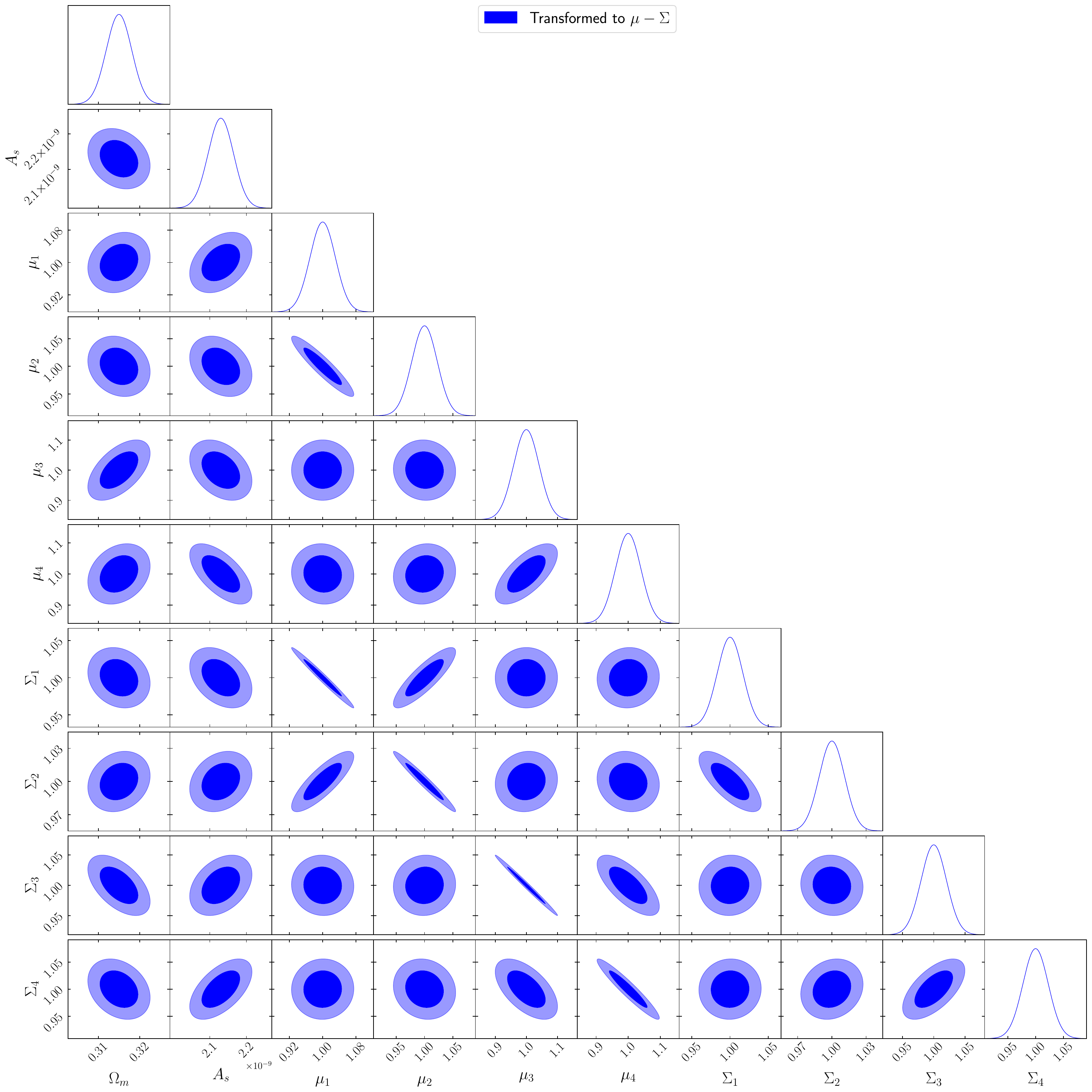}
    \caption{The analogous version of fig.~\ref{fig:quasi_cut} with the coordinate transformation $(\mu, \eta) \rightarrow (\mu, \Sigma)$ where $\Sigma = \mu(1+\eta)$. We see that the overall constraining power on $\mu$ goes down but the constraining power on $\Sigma$ increases relative to that on $\eta$. }
    \label{fig:SigmaTransformation}
\end{figure}

We now turn to the subject of alternative parameterisations. We show in fig.~\ref{fig:SigmaTransformation} the Fisher contours associated to the $\mu-\Sigma$ parameterisation. We explicitly see here the $\mu-\Sigma$ degeneracy discussed in sec.~\ref{subsec:degeneracies} and the overall change in the constraining power on the MG parameters relative to the standard $\mu-\eta$ case considered in the main text. In the $\mu-\Sigma$ case, the transformation homogenises the constraints in the MG parameter space, i.e., the relative error on $\mu$ is comparable to that of $\Sigma$, as opposed to $\mu$ and $\eta$ (see sec.~\ref{subsec:degeneracies} for more details). In the main text, we made the choice of presenting our results using the $\mu-\eta$ parameterisation, since it leads to better constraints on $\mu$, which is the parameter that requires our non-linear recipe and this is varied in the $N$-body simulations used to validate our non-linear recipe. We note that the specific probe combination being considered is particularly relevant for this choice. Galaxy clustering constrains $\mu$ alone, while weak-lensing directly constrains $\Sigma = \mu(1 + \eta)$. The nature of the data in question (spectroscopic/photometric) is also relevant to the choice of parameterisation, particularly because spectroscopic surveys will not be able to effectively constrain $\eta/\Sigma$. We conclude that these considerations make it difficult to make an overall blanket statement about which choice is better in general. Instead, we suggest that these results indicate what one can expect for a typical photometric survey, and may be used as a guide to make an informed decision depending on the goals of the analysis one is interested in.

\end{document}